# Institutions and China's Comparative Development

Paul Minard
Cégep Heritage College
Saint Paul University
pminard@cegep-heritage.qc.ca

**Abstract.** Robust assessment of the institutionalist account of comparative development is hampered by problems of omitted variable bias and reverse causation, since institutional quality is not randomly assigned with respect to geographic and human capital endowments. A recent series of papers has applied spatial regression discontinuity designs to estimate the impact of institutions on incomes at international borders, drawing inference from the abrupt discontinuity in governance at borders, whereas other determinants of income vary smoothly across borders. I extend this literature by assessing the importance of sub-national variation in institutional quality at provincial borders in China. Employing nighttime lights emissions as a proxy for income, across multiple specifications I find no evidence in favour of an institutionalist account of the comparative development of China's provinces.

August 14, 2019

# I.    Introduction

Recent decades have seen the elaboration of an institutionalist paradigm in the study of comparative economic development.  This paradigm emphasizes that institutions, "the humanly devised constraints that shape human interaction" (North, 1990, p.3), structure the incentives shaping economic exchange.  From a growth perspective, good institutions are those which reduce transaction costs and secure property rights, which will encourage investments in human and physical capital and facilitate economic exchange.  Poor institutions are those which limit access to secure property rights to elite allies of the state.   In the absence of secure property rights, high transaction costs deter investments and impose costs on economic agents who must attempt secure these rights privately.  As such, the institutions of formal governance are modeled as the fundamental driver of long-run economic performance.

Intuitively, differential growth rates in the wake of the separation of the two Koreas and the two Germanys suggest that the institutionalist account is plausible (Acemoglu and Robinson, 2005).  Robust empirical assessment of this paradigm, however, has proven challenging given the dangers of confounding and reverse causation.  While cross-country analyses undertaken in the 1990s uncovered associations between numerous measures of institutional quality and economic performance (e.g., Knack and Keefer, 1995; Hall and Jones, 1999), it may be that since quality institutions like property rights protection are costly to enforce, richer societies can afford better institutions, which would reverse the causality underlying the observed association. Clark (2007), for instance, suggests that rather than good institutions facilitating trade by protecting property and reducing uncertainty in exchange, increased opportunities for trade or demand for the products of a region can increase demand for trade-supporting institutions.  Rather than being a determinant of trade volumes, good institutions may therefore be a response to increased opportunities for trade owing to demand shocks, reductions in transportation costs owing to technological change, and the like.  These early findings may also have been subject to omitted variable bias, as additional factors like geography, human capital endowments or culture may drive both institutional quality and economic performance.[1]  Indeed, it is a stylized fact that richer countries, in addition to having more highly rated institutions, tend to have higher human capital levels than do poorer countries, and in many cases are more favourably situated in a climactic and geographic sense.

---

[1] For instance, Glaeser et al (2004) revisit standard cross-country investigations and find that while average constraints on government

More recently, economists have sought to leverage plausibly exogenous historical sources of variation in institutional quality in institutional quality to estimate the causal effect of institutions on economic growth in an instrumental variables framework (e.g., Acemoglu, Johnson and Robinson 2001, 2002). However, Glaeser et al (2004) question whether these papers have adequately satisfied the exclusion restriction on which the estimation approach relies, and moreover propose that a human capital-centric interpretation of these results is more plausible. Indeed, as discussed below, literatures emphasizing human capital, geo-climactic, and ethnocultural determinants of development contend that the contend that institutionalist accounts are subject to confounding.

This paper contributes to an emerging literature which employs spatial regression discontinuity designs (RDDs) to estimate the causal effect of institutional quality on economic development at borders. RDDs enable robust inference of causal effects while relying on less restrictive assumptions than alternative approaches, and as such have been validated as a more robust method for establishing causal inference than other forms of natural experiments (Cook, Shadish and Wong, 2008). Spatial RDDs are ideally suited to the study of the effect of government-backed institutions on income where the institutions under which people live vary discontinuously at a defined administrative boundary, whereas other determinants of income vary smoothly across these borders. Recent applications of a spatial RDD approach to the study of institutions include Michalopolos and Papaioannou (2014) who use a spatial RDD to examine whether incomes vary discontinuously at national borders in Africa, finding no effect once ethnicity has been controlled for. Dell (2010), however, finds evidence of a legacy from extractive institutions hundreds of years ago on current incomes in Peru. In a cross-national study, Pinkovskiy (2017) finds abrupt differences in growth rates at international borders, implying the relevance of national institutions, or national-level cultural differences, to incomes in border regions where geographic and climactic conditions are otherwise comparable.

---

across a period of years is associated with economic growth in the same period, the initial level of constraint is not predictive of subsequent growth. Initial human capital levels, however, are predictive of future economic performance. Moreover, while there is a robust association between assessments of institutional quality and growth, there is no association between constitutional constraints on government power and growth.

This methodology has not yet been extended to estimate the effect of subnational institutions on comparative development within countries. In this paper I employ the spatial RDD approach to assess the relevance of institutions as an explanation for patterns in comparative development among Chinese provinces. China's size, 31 provinces, relative cultural and ethnic homogeneity, regional income disparities and institutional diversity make it an ideal setting for the employment of a spatial RDD to test an institutionalist account of subnational comparative development. I use World Bank (2008) assessments of province-level institutional quality to identify higher and lower quality institutions among 68 pairs of provinces, and test for discontinuities at provincial borders as one moves from a province with relatively lower quality institutions to one with relatively higher quality institutions. If province-level institutional differences are an important driver of comparative development within China, discontinuities in income should be evident at provincial borders.[2]

In the spirit of conventional RDDs, inference is drawn from regions in the immediate vicinity (50km at the greatest extent) of either side of a provincial border, where the identifying assumption of comparability in observed and unobserved covariates on either side of the border is most easily sustained. In the results section below, I first demonstrate that a series potential confounds that may also affect income levels (precipitation, elevation, population and public goods provision) vary smoothly across borders, substantiating the assumption of comparability underlying the spatial RDD (Keele and Titiunik, 2016).

Though China is ethnically relatively homogeneous, within Chinese territory there is considerable diversity in dialects within the Sino-Tibetan phylum, which may reflect cultural differences that are relevant for comparative development. Figures 1a and 1b represent the traditional homelands of Chinese ethnic and linguistic groups, respectively. Where provincial borders track these cultural borders, the spatial RDD will not enable us to attribute a discontinuity in income to a discontinuity in institutional quality, raising the spectre of multiple treatments and violating the assumption of RDDs. To guard against the danger of noncomparability, in the main specification I therefore derive estimates from within-dialect group variation in institutional quality on the basis of the Language Atlas of China (Lavely and Berman, 2012).[3]  In subsequent analyses, I present separate RDD estimates at each of 22 provincial borders representing a putatively large discontinuity in institutional quality.  Proxying for income using nighttime lights emissions, in the main analysis and across each of these individual borders, I find no discontinuity in incomes.  The finding is robust to three operationalizations of nighttime lights emissions and multiple RDD specifications.  These findings suggest that institutional diversity is not the primary driver of China's subnational comparative development.

---

[2] I show below that while the World Bank assessments were conducted in provincial capitals, there is compelling evidence that the resulting rankings reflect province-level variation in institutional quality.

**Insert Figures 1.a) and 1.b) about here**

**Insert Figure 2 about here**

In the next section, I discuss contending paradigms of comparative development emerging in the recent literature which highlight the need for a robust methodological approach to the study of the impact of institutions.  I also provide a brief assessment of China's institutional diversity, and describe existing evidence on the explanatory power of institutional variation with respect to interprovincial income inequality.  Section III describes the data employed and the paper's identification strategy.  Spatial RDD estimates of the effect of institutional quality on incomes at a series of provincial borders are presented in Section IV.  Section V concludes with a consideration of the implications of these findings for the study of comparative development in China and beyond.

## II.      Literature review

*Contending accounts*

Among skeptics of the institutionalist account of comparative development are those who argue for the primacy of geographic, topographic and climatic factors.  The "geo-centric" account emphasizes, for instance that challenging topography, low proximity to global markets and inhospitable climates cannot be overcome with quality institutions.  Hence for instance, land-locked, mountainous Bolivia, despite having institutions rated better than Vietnam's, has had difficulty attracting foreign investment without the advantages of Vietnam's extensive coastline, deep water ports and proximity to other Asian growth centres (Sachs, 2012). Sachs (2003, 2012), also notes that various measures of economic performance are correlated with climate zone and disease ecology across countries. Even where an association between institutional quality and economic performance can be demonstrated, in the geocentric account, favourable geography generates trade opportunities, which increases the value of local endowments, which in turn increases demand for secure property rights. The costs of securing property rights are then worth bearing, whether privately or publicly. The geo-centric account does not go so far as to say that institutional quality in the form, say, of property rights protection, has no role to play in comparative development. If two locales are equal in their geographic endowments, these advantages will leveraged most effectively in the locale with more secure property rights. However, the magnitudes of these differences must be assessed where geography is a constant, and are, in this view, likely to be small relative to geographic drivers.

---

[3] The intuition behind this approach is represented in Figure 2, which shows the Gansu-Sichuan border.  Estimates of the impact of Sichuan's higher-quality institutions are derived separately within the regions traditionally occupied by speakers of Mandarin-related dialects (in red) and regions traditionally home to speakers of Tibeto-Burman dialects (in green).

Skepticism of the institutionalist account has also emerged in a series of papers investigating the historical, cultural, and possibly genetic "deep roots" of comparative development. Consistent with a theory of the persistence of initial advantages in technology adoption, Comin, Easterly and Gong (2010) show that the level of technology adoption in the year 1500 AD is a strong predictor of technology adoption and per capita income today. There is also robust evidence of technological persistence from as early as 1000 BC. Putterman and Weil (2010) find that the share of a country's population in the year 2000 which is descended from populations which in 1500 were living in locales that had longer histories of functioning states and earlier transitions to agriculture is predictive of contemporary incomes. In a comprehensive summary of this literature, Spolaore and Wacziarg (2013) note that both cultural and biological transmission of traits favourable to economic development are possible, though whether these traits impact development directly, or indirectly as barriers to the diffusion of technology across populations, remains unclear (see also Spolaore and Wacziarg, 2009). In a paper linking the bio-cultural and geo-centric accounts, Galor and Ozak (2016) provide evidence that agro-climatic conditions favourable to the early adoption of agriculture gave rise to selection and/or learning effects that have been transmitted to modern populations. Having ancestors exposed relatively early to agriculture predicts a long-term orientation in one's time preference, which is favourable for economic development to the extent that it predicts a higher savings rate and a long-term orientation to investment decisions. A related literature also shows that levels of trust and social capital within populations, irrespective of institutional quality, is predictive of the spatial distribution of economic development within countries (e.g., Tabellini, 2010).

*Subnational institutional diversity in China*

Several stylized facts make China an interesting case for assessment of the institutionalist account of development via a spatial RDD. Given China's remarkable growth in recent decades and the absence of evidently inclusive nationwide institutions in the Northian sense, some scholars have viewed China's growth as a puzzle for institutionalists (Xu, 2011; Lu, Png and Tao, 2013). At the same time, other scholars have credited unique features of China's decentralized approach to economic policymaking as being essential institutional foundations of China's rapid growth (Montinola, Qian and Weingast, 1995; Weingast, 1995; Qian and Weingast, 1997). One point of agreement among scholars is that while not nominally a federal state, economic policymaking in China is highly decentralized (Xu, 2011; Birney, 2014; Fuller, 2007). According to one estimate, approximately 74% of all government spending in China is undertaken at the local (i.e., provincial and below) level (Wong, 2007). Moreover, the lines of authority between central government ministries and provincial governments, and the degree of government control over public service units (e.g., schools and health facilities) are legal grey areas. Local officials, then, have "immense discretion" (Birney, 2014, p.1) over governance, including fiscal policy, the provision of public services, the making and enforcement of laws, the allocation of firm finance, contract enforcement and the protection of property rights (Xu, 2011). Given this discretion, a system has emerged in which the regulatory environment is "fractured" (Pearson, 2007), resulting in policy heterogeneity (Poncet and Barthelemy, 2008). This decentralization has led to the rise of "locally distinctive models of regional development" at the provincial level (Wei, 2007, 19). Hence scholars identify, for instance, the "Jiangsu model" (Xianping, 2004), "Guangdong model" (Xu and Zhang, 2009), or "Zhejiang model" (Ye and Wei, 2005) of development, which are characterized in part by diverse approaches to property rights and ownership structures among firms. Lin and Liu (2006) argue that China's regional income disparities are "endogenously determined" by differing provincial development strategies, with interior provinces continuing to pursue misguided economic policies.

This province-level variation in institutions is reflected in several quantitative assessments. To take two components of World Bank measures of the cost to firms of securing their property rights, the time required to resolve a sale of goods dispute ranges from 112 days in Nanjing (Jiangsu province) to 540 days in Changchun (Jilin province), while the cost as a percentage of the value of the claim varies from 9% in Shanghai to 41.8% in nearby Hefei (Anhui province). In a 2004 World Bank survey that asked over 12,400 Chinese firms to assess the percentage of cases in which legal contracts or properties had been upheld, provincial average responses ranged from 41% in Gansu province to 83% in Zhejiang. The World Bank (2008) has documented substantial variation in proxies for corruption across Chinese provinces. There is a broad consensus across various assessments as to the provinces in which institutional quality is relatively high. The correlation in provincial rankings between two leading indexes (the World Bank's measure of property rights referenced above and employed further below, and the National Economic Research Institute's marketization index), is 0.72.

Per capita income also varies dramatically throughout China, incomes in the richest provinces being on par with those in Eastern Europe, while India is a better comparison for the poorest provinces. Indeed, China's substantial population of migrant workers is a testament to the existence of wage gaps across the country. Since institutional quality also varies substantially across provinces, it is tempting to conclude that institutions are driving economic performance in China, and thus account for a substantial share of regional inequality. Figure 3 presents the association between World Bank provincial ranks in property rights security and per capita GDP in 2008. The $r^2$ in this simple regression indicates that 50% of the sample variance in GDP is accounted for by variation in the ranking of institutional quality.

**Insert Figure 3 about here**

Institutionalist accounts in the Chinese case have found higher firm investment rates when property rights are perceived to be secure (Cull and Xu, 2005; Hallward-Driemeier, Wallsten and Xu, 2006). Property rights security also appears to positively impact firm research and development spending (Lin, Lin and Song, 2010) and productivity (Lu, Png and Tao, 2013). The location choice of US multinationals within China is also correlated with property rights security (Du, Lu and Tao, 2008).

Much of the institutionalist literature on China's comparative development has presumed that institutional quality is exogenous to local geographic and human capital endowments. Reporting on a World Bank survey of investment climate across 120 Chinese cities, for instance, Mako (2006) states that, "cities in the bottom quintile of 'government effectiveness' could expect near-term gains of 25-35 percentage points in firm productivity and 15-25 percentage points in foreign ownership" (Mako, 2006, 6) were governance improved to the level found in China's best governed provinces. Dollar et al (2004) similarly claim that an average Chinese city could see firm productivity increases of 45 percent by improving governance to the level in cities in the 90[th] percentile.

If institutional quality is endogenous, however, these estimates are unreliable. It is notable, for instance, that provinces rated as having better institutions tend to be those along China's coast, and which thereby enjoy several geographic advantages over their inland neighbours. Bao et al (2002) note that as one moves inland, China's topography becomes increasingly mountainous, implying high transportation costs and a resulting barrier to trade and technology diffusion. This is true of precipitation levels, too, which is particularly important in China, where in the years under study about 37% of the population was engaged in agriculture (World Bank, 2017), and the legacy of agricultural productivity improvements in the early reform years has been important for subsequent industrial development. Though the interior provinces are richer in mineral resources, these are costly to extract and transport owing to geography. The wealthier coastal provinces are also more easily connected to world markets, and have deeper familial ties with the more economically and technologically advanced societies in Taiwan and Hong Kong. According to a geocentric reading of the evidence, then, once the post-Mao reforms were enacted and China opened again to the outside world, the inherent advantages of eastern provinces could be leveraged, and differential growth rates driven by geography followed.

Research applying "deep roots" theories to China is limited. Bai and Kung (2011) assess the relative genetic distance to Taiwan for all pairs of Chinese provinces, and find that since the reopening of China-Taiwan travel in 1987, relative (but not absolute) genetic distance from Taiwan has been predictive of higher per capita GDP. On the assumption that visitors to the mainland from more-developed Taiwan were more likely to visit and trade with their relatives, this finding implies that traits favourable to development diffuse through communication and economic exchange, rather than directly through biological inheritance. Bai and Kung (2015) report further evidence that biological and cultural differences within China have historically been barriers to technology diffusion, which questions the primacy of institutional quality as the most binding constraint on income convergence.

## III.    Methodology

*Regression discontinuity designs*

China provides an ideal setting for the application of a spatial RDD to assess an institutionalist account of comparative subnational development. China has 31 provinces[4], among which there are 68 border pairs. We are provided, then, with numerous quasi-experiments in institutional quality at provincial borders, as one moves from a province of relatively lower to relatively higher institutional quality. Despite the conventional summation of centre-local relations implied by the proverb, "the mountain is high, and the emperor far away", the penetration of local government into the lives of citizens in even remote areas has a long history in China. For instance, Chen and Lan (2017) document that during the agricultural collectivization drive of the mid-1950s, government had managed to enforce collectivization upon 96% of the rural population. This indicates the reach of provincial institutions of governance is not limited to capital cities. China is also a relatively homogenous country in terms of culture and ethnicity, and while larger minority groups are concentrated in some broad regions of the country, the rough boundaries of these groups do not precisely follow provincial borders. As noted above, I nevertheless take the cautious approach of deriving main estimates from within-dialect group variation in institutional quality. Similarly, while China is geographically diverse, provincial borders do not track abrupt geographic or climactic discontinuities, nor discontinuities in infrastructure, as I am able to show below. Historians are also of the broad view that the precise location of provincial borders is arbitrary with respect to geographic, cultural and economic territories, and that this has been true since at least the Ming Dynasty (Fitzgerald, 2002).

---

[4] This count includes 22 provinces, five autonomous regions and four province-level municipalities.

In a straightforward RDD, observations are assigned to a treatment condition, *d=1*, based on the value of a running variable, *c*. RDDs are appropriate where *d* is a discontinuous function of the value of c at some precisely defined cutoff, *c\**. Thus for observations $>=c*$, *d=1*, whereas for observations $<c*$, *d=0*. Robust inference of the effect of treatment on the outcome variable depends on the assumption that omitted variables vary continuously around c*, whereas treatment varies discontinuously. Spatial RDD studies of institutions treat geographic location as the running variable, *c*, and an observation's location on, in this case, the high quality institution side of a border as determinative of treatment. Identification in a spatial RDD relies on the assumption that local determinants of economic performance vary continuously at the border, whereas governance varies discontinuously. Thus while climate, topography, distance to population centres or export markets vary smoothly across the border, it is as if otherwise similar locations have been exposed in a quasi-random fashion to different institutions. If the assumption of smooth variation across the border in these covariates can be sustained, a spatial RDD provides a unique method for an unbiased assessment of the effect of good institutions on economic performance at borders. Moreover, by limiting our analysis to territories within a narrow band along provincial borders, we can be confident that local conditions are not determinative of province-level institutions, and hence that provincial institutions are exogenous to local conditions.

It should be emphasized that the local treatment effect estimated via a spatial RDD is based on all discontinuities at the border. I am able to show below that a series of geographic and climactic variables vary smoothly across provincial borders, as do population levels. Yet as administrative borders mark changes in governing authorities, any discontinuities in income would reflect the sum effects of all government policy, including, for example, public goods provision.  I therefore establish comparability in terms of a proxy for public goods provision, road density, to further ensure that any observed discontinuities reflect the quality of governance rather than all government activity per se.

*Data and variables*

I proxy for income in border regions using nighttime lights emissions.  These emissions are recorded by the National Oceanic and Atmospheric Association (NOAA)'s Defense-Meteorological Satellite Program - Operational Linescan System (DSMP-OLS).  Though these satellites were initially launched to gather data on cloud cover to improve weather forecasting, they also record light emissions, and are sensitive enough to detect the luminosity of street lighting (Keola, Anderson and Hall, 2015).  The DSPM-OLS satellites orbit the earth daily, recording light emissions between 20:00 and 22:00 local time at a resolution of approximately $1km^2$ at the equator.  Each of these $1km^2$ pixels is coded with a digital number representing luminosity on a scale ranging from 0 to 63.  Worldwide pixel-level datasets are released annually by the NOAA which average the recorded luminosity values recorded on cloud-free nights during a calendar year.[5]

Numerous studies have validated nighttime lights emissions as a proxy for income (e.g., Henderson et al., 2012; Min and Gaba, 2014; Klemens et al., 2015).  Henderson et al (2012) estimate the elasticity of luminosity growth over time and traditional income data in countries with robust statistical data, finding that the long run growth rate in luminosity is approximately equal to the long run rate of income growth as measured more traditionally.  Nighttime lights are a particularly useful proxy for income in a developing country context, where official government data are often unreliable (Jerven, 2013) and household survey data, when available, often fails to accord with official statistics (Ravallion, 2003).  These are important concerns in the Chinese context for two reasons. Firstly, given the incentives of Chinese officials to overstate economic growth in their locales (Edin, 2003), the reliability of official Chinese economic data has been questioned (e.g., Chow, 2006; Balding, 2013; Koch-Weser, 2013). Moreover, many Chinese, especially in rural areas, earn income in the informal sector and are paid untaxed wages in cash that go unrecorded by government. It is also well documented that China has a large population of migrant workers who work and reside away from their official place of residence as recorded by the state. Even government income statistics based on probability sampling of local populations sample only legally documented residents. This is thought to introduce substantial error into per-capita income estimates, the direction of which varies across provinces, some being net importers of labour, and others net exporters (Li and Gibson, 2013). Indeed, the only existing application of nighttime lights data in the Chinese case finds, contrary to government data, no broad reduction in rural poverty from 2005 to 2010 (Almas, Johnsen and Kotsadam, 2014). It is also worth noting that once state ownership of firms is taken into account, China's GDP statistics are driven substantially by government investment, rather than the consumption of citizens (Lardy, 2006). Given concerns about unproductive investment and corruption, a robust, apolitical measure of the consumption of ordinary citizens is particularly desirable in the Chinese case. Finally, the nighttime lights data can be processed to construct treatment and control regions of the desired bandwidth by the researcher, including areas adjacent to the border, thereby enabling robust inference via the spatial RDD.

---

[5] I employ the NOAA's stable-nighttime lights dataset.  To access the data, and for a description of other available datasets which have not been pre-processed to remove ephemeral lights as caused by forest fires, the aurora borealis, and the like, see http://ngdc.noaa.gov/eog/dmsp.html.  For further descriptions of the data and download process, see Noor et al (2008), Chen and Nordhaus (2011), Henderson, Storeygard and Weil (2012), and Addison and Stewart (2015).  Sources for all variables are identified in Table 12.

To assess the effect of institutions on luminosity, I identify 68 inter-provincial borders in China[6]. I identify one of each pair of provinces as having more highly rated institutions. For each pair of provinces, the treatment variable, *high quality*, takes the value of 1 for the province with the higher-rated institutional quality, and 0 for the province with the lower-rated institutional quality. Institutional quality is proxied using the World Bank's *Doing Business in China* (2008), which ranked Chinese provinces according to the ease of doing business in the province. One component of this ranking is a broad assessment of the ease of contract enforcement. I take this as a proxy for property rights, an essential feature of high quality institutions.[7] As this

---

[6] Hainan, China's only island province, is excluded from the analysis. The Hong Kong and Macau special administrative regions (SARs) are also excluded. Borders are defined using GADM databases of Global Administrative Areas. See www.gadm.org.
[7] For a detailed description of the construction of this measure, see World Bank documentation at

measure of de facto property rights protection was assessed by the World Bank in 2008, I take mean luminosity across 2009 and 2010 as the dependent variable. Using what amounts to a lagged measure of institutional quality helps to further guard against the possibility of reverse causation (Michalopoulos and Papaioannou, 2014).

While the World Bank report presents assessments of province-level institutional quality, their rankings are based on surveys conducted in provincial capital cities only. Though the variation in contract enforcement uncovered by these surveys is substantial, by itself their report does not enable us to conclude with confidence that these differences across provincial capitals are in fact reflective of province-wide differences. That is, though the capital city of Guangdong (Guangzhou) may offer better contract enforcement than the capital of neighbouring Hunan (Changsha) the World Bank report does not document whether the purported Guangdong-Hunan difference in institutional quality is meaningful at the border between the provinces. This paper's substantive conclusion rests on the assumption that these inter-provincial differences in institutional quality are indeed province-level differences which vary discontinuously at borders, and not unwarranted generalizations on the basis of varied provincial capitals that sit as islands in a sea of institutional sameness. I begin the results section below with evidence in support of this assumption.

To define the geographic scope of analysis around each border segment, I first use ArcGIS 10.4 to create "buffer zones" of 50 kilometres on either side of each border which are then divided into cells of 0.05 by 0.05 decimal degrees. The modal cell has an area of approximately 30 km$^2$. Our main dependent variable, *luminosity*, can be thought of as the natural logarithm of the total light emissions within a cell on a typical night in the years 2009 and 2010.[8] In addition to this measure of the intensiveness of luminosity, I also assess the effect of treatment on an extensive measure by constructing a cell-level dummy variable, *lit*, which takes the value of 1 if any luminosity is present in the cell. ArcGIS tags the centroid of each cell by its longitude and latitude coordinates, so that its precise location is known. Limiting the analysis to, at the largest extent, a 50km

http://www.doingbusiness.org/data/exploreeconomies/china/enforcing-contracts. This measure is not available for Tibet, and so borders between Tibet and its neighbours are excluded from the analysis.
[8] Given nearly 58% bottom-coding, I add 0.01 to the raw value of the sum of emissions from pixels within a cell, and take the logarithm of this value. See Michalopoulos and Papaioannou (2014).

buffer on either side of the border helps to reduce the possibility of confounding, on the intuition that unobserved determinants of growth in each province are more similar as one moves toward the border. Where observed covariates are concerned, this is documented empirically.

This approach is illustrated in Figure 4a, using the border between Heilongjiang and Jilin provinces as an example. The bright green line running roughly from the northwest to the southeast represents the border. The grid overlaying the image is a "fishnet" created in ArcGIS which defines the unit of analysis, the cell. The extent of the fishnet indicates the 50km "buffer zone" around the border. Cells to the north of the border are in Heilongjiang, and those to the south in Jilin. The background colour indicates pixel-level luminosity, lighter shades reflecting higher light emissions. Figure 4b is a higher resolution image at the same border.

**Insert Figure 4a about here**

**Insert Figure 4b about here**

In an effort to ensure the comparability of cultural groups on either side of a border between high and low quality institution provinces, in the main specification I derive estimates of the impact of institutions on incomes from within dialect-group comparisons. I use the Language Atlas of China (Lavely, 2000; Lavely and Berman, 2012) to assign each cell to a dialect group (see Figure 1.b). Virtually all of mainland China is coded as within the Sino-Tibetan phylum. Within the Sinitic stock, Lavely (2000) sorts dialects into groups, "within which languages are assumed to be mutually intelligible" (Lavely, 2000, p.2), whereas dialects across these groups are mutually unintelligible. In Lavely's coding, this results in the following groups: Mandarin group, Jin group, Gan group, Xiang group, Min group, Yue group, Hakka group, Hui group and a residual category composed of dialects spoken in only one or two counties. The Language Atlas also identifies sub-groups which code dialects more narrowly, including those outside the Sinitic stock. Given that Lavely's (2000) original focus was on intra-Han dialect variation, I modify his coding by adding the following sub-groups to "group" status: Miao-Yao, Wu, and Yue (Cantonese). I also code Mongolian and Tibeto-Burman dialects as distinct groups. This coding results in 19 dialect groups. After dropping groups which cover less than one percent of cells in border regions, I am left with the eight groups noted in Table 1. Note that 67% of border cells are within the Mandarin group.

<div align="center">**Insert Table 1 about here**</div>

*Estimation*

Assignment of a cell to the treatment condition of relatively higher quality institutions in this context is a deterministic, discontinuous function of geographic location (latitude and longitude). Robust inference in a spatial RDD requires isolating the discontinuous border effect of geographic location on the outcome from the smooth effect of location on the outcome. There are two basic approaches to modeling this smooth effect. One approach is to employ nonparametric techniques, such as local linear regressions. This approach limits the analysis to observations within a specified bandwidth around the border. The intuition behind local linear regression is that by focusing on observations near the border, we reduce the danger of bias as we do not need to make assumptions about the functional form of the effect of geography on the outcome variable. The limitation of local linear regressions is that while bias is reduced in limiting the geographic scope of the analysis, this means drawing inference from fewer observations. A second approach is to model the effect of geography directly in the context of a parametric regression including polynomials for geographic location. Under this approach, it is important to model these effects appropriately, otherwise, for instance, a nonlinearity in the location-outcome relationship near the border could be mistaken for a discontinuous treatment effect. This could introduce bias into estimated treatment effects. The trade-off between the parametric and nonparametric approaches, then, is between bias and variability.

The RDD literature recommends nonparametric techniques to avoid the possibility of bias in modeling choices, provided there are sufficient observations near the cutoff to minimize variance (Imbens and Lemieux, 2008; Dell, 2010). Moreover, Keele and Titiunik (2015) emphasize that spatial correlation can be dealt with effectively using nonparametric estimation. Since the nighttime lights data are available at very high resolutions, we have a large dataset at most borders. I therefore focus on nonparametric local linear regressions as my primary specification, as recommended by Athey and Imbens (2017).

Estimating local linear regressions requires selection of a bandwidth (i.e., the extent of a buffer zone) around the discontinuity, and a bin width within which average values will be calculated. Rather than rely on an ad hoc approach to bandwidth and bin selection, I employ the data-driven approach described in Calonico, Cattaneo and Titiunik (2014) and Calonico, Cattaneo, Farrell and Titiunik (2017), which uses the underlying structure of the data to select an optimal bandwidth and bin width for estimation, and which allows the inclusion of covariates[9]. In our primary specification with dialect group fixed effects, the basic functional form estimated at each border can be written as:

$$y_{ijp} = \alpha_0 + \beta IQ_p + f(D_{ijp}) + \alpha_j + \varepsilon_{ijp} \tag{1}$$

where $y$ is the outcome variable in cell $i$ in language group $j$, province $p$. A cell's location on the high-quality institution side of the border is denoted where $IQ$=1. On the low quality side, $IQ$=0. $\alpha_j$ represents a series of dialect group dummies. $\beta$ is the coefficient of interest, the estimated ceteris paribus effect of high quality institutions on luminosity. Depending on the specification, $f(D_{ijp})$ enters as either linear and quadratic RDD polynomials, which control for the smooth effects of geography on the outcome, and which allow these effects to vary on either side of the border. These are a polynomials of one-dimensional distance to the border, with a discontinuity at zero. Where the identifying assumption of no discontinuities in observed covariates at the cutoff is sustainable, opinion in the RDD literature on the desirability of the inclusion of covariates is divided (see the discussion in Calonico et al., 2018). Given the absence of discontinuities in control variables at borders as per the estimates shown below, and given the large number of cells at each border, the main specification does not include covariates. However, in the interest of transparency, in a secondary specification I also include a series of cell-level covariates that may also affect the outcome variable, $X_{ijp}$. These include mean elevation within the cell, the sum of annual precipitation within the cell, the logarithm of population within the cell according to the 2005 population census, the distance to the nearest road in kilometres, and the logarithm of the cell's area. Cells with no population are omitted from the analysis. Controlling for population density also ensures that the effect of institutions beyond the effect on population density is assessed (Michalopoulous and Papaioannou, 2014)[10]. Controlling for distance to the nearest road helps guard against the multiple treatments concern, in that the measured effect of institutions will be net of the effect of public goods provision.

---

[9] Implemented using the Stata command "rdrobust". The accompanying package, "rdwselect", selects bandwidth that minimizes mean squared error. Standard error estimates in all specifications are reported using the cluster robust variance estimators. See Calonico et al (2017) for more details. To download this package and accompanying documentation, visit https://sites.google.com/site/rdpackages/rdrobust.

In addition to this main specification with dialect group fixed effects, I also undertake spatial RDDs at a series of individual borders. Differences in ranking between bordering provinces of course vary, from a low of one ranking spot (between Shaanxi and Chongqing) to a high of 26 (between Guangdong and Hunan). The mean difference in rank among 68 border pairs is 8.79, with a standard deviation of 6.51. Whatever discontinuity exists between the former pair of provinces in institutional quality, the institutionalist argument is more fairly assessed where the gap in institutional quality is thought to be large. I therefore estimate separate regressions using a modified version of the above model without dialect group fixed effects for each border segment where the gap in World Bank ratings is at least one sample standard deviation (and thus, is seven ranking spots or larger). This leaves 22 border pairs. [11] This exercise can be interpreted as a robustness check on our main, within-dialect groups fixed effect analysis, insofar as it substantiates the consistency of our main finding across a large number of particular borders.

## IV.    Results

*Governance discontinuities at provincial borders*

---

[10] An alternative approach to controlling for population would be to deflate our dependent variable by population. As noted by Michalopoulos and Papaioannou (2017), however, non-linearities in the elasticity between population and luminosity could lead to unpredictable biases. Population therefore enters the regression on the right hand side in this case.

[11] As noted below, given the large number of border pairs, to be conservative about the avoidance of bias owing to non-comparability, I omit from the analysis one border pair at which there is some evidence of a discontinuity in road density in a narrow band around the border. As data for numerous covariates are available, following Keele and Titiunik's (2016) terminology, I take the conservative position that treatment and control cells at each border are "comparable", and can thus be modeled through the inclusion of controls in regressions, rather than assume treatment and control regions are perfect counterfactuals for each other (i.e., differences in covariates are "ignorable").

I first present evidence that the World Bank's rankings of governance quality are reflective of province-wide differences in governance, and thus that institutional quality differs at provincial borders, and not merely across provincial capitals. I present two lines of evidence from different data sources. In the first I employ data from two World Bank surveys of Chinese firms, undertaken in 2004 and 2006. In each survey, 100 firms were sampled in each of the same 120 cities. Each survey included a module on the firm's experience interacting with government and the legal system. Five questions in particular are germane to firms' perceptions of institutional quality. These are as follows:

1. "Amongst the commercial or other disputes that your company was involved with, what has been the likelihood (in terms of percentage) that your company's contractual and property rights (including enforcement) are protected?" World Bank (2006) presents city-level aggregates, standardized in order to construct an index.
2. Travel and entertainment costs, relative to sales (World Bank, 2006). City-level aggregates computed by World Bank.
3. "How many days does the GM or vice GM spend on government assignments and communications per month? (Government agencies include Tax Administration, Customs, Labor Bureau, Registration Bureau, etc.; assignments refer to handling the relationship with government workers, consolidating and submitting various reports or statements, etc.)" (World Bank, 2004). Firms were presented with the following ordinal response categories: (1)1 day (2)2-3 days (3)4-5 days (4)6-8 days (5)9-12 days (6)13-16 days (7)17-20 days (8) > 21 days. I compute city-level mean responses.
4. "In the case of commercial disputes with suppliers, clients or subsidiaries in your province, how much confidence do you have that the disputes will be settled with justice by the local legal system" (World Bank, 2004). Firms responded in percentage terms. I compute city-level mean responses.
5. "In commercial or other legal disputes, what percent of cases were your company's legal contracts or properties protected" (World Bank, 2004). Firms responded in percentage terms. I compute city-level mean responses.

I assemble a city-level dataset consisting of these variables, the province in which the city is located, and the city's geographic coordinates. I then populate a list of dyads consisting of city pairs, and calculate the absolute difference in city average responses to each of these five institutional variables. The dataset is then limited so that for each city, its pair is the closest city in either its own province or in a province with which it shares a border. A limiting distance of 150km between city pairs is imposed. The resulting dataset consists of 150 pairs of neighbouring cities which are distinguished as being "within-province" neighbours or "across-province" neighbours. If the assumption that variation in institutional quality across China is indeed province-level is sustainable, we would expect smaller differences in perceived institutional quality among neighbouring firms within the same province than among neighbouring firms separated by provincial borders. Figure 5 compares mean differences among pairs of across-province neighbours with those among within-province neighbours. For each of our five measures of institutional quality, larger differences are found across provinces than within provinces. Recall that we have imposed an upper limit of 150km on how distant a city can be from its pair. Among these neighbouring cities, separation by a provincial border has a predictable effect of increasing difference in mean responses to these five survey questions. Though the effects are not dramatic in magnitude, the consistency of this effect argues in favour of province-level variation in institutional quality.

**Insert Figure 5 about here**

Our second line of evidence comes from prefecture-level data on the share of workers employed by private enterprises. Private enterprises remain a politically sensitive business form in China to the extent that they threaten the rents enjoyed by state-owned enterprises and, though they generate tax revenues, are not effective instruments of state policy. Household enterprises were recognized as a legitimate commercial form only in 1981, and private enterprises in 1988 (Young, 1989). Private enterprises have been frequent targets of expropriation by government (Chen et al, 2011; Huang, 2008), and local governments have had, "considerable discretion in establishing the nature and size of fees to be paid by enterprises under their jurisdictions" (Choi, 2009, p.80). The credit constraints faced by private enterprises are also well-documented (e.g., Poncet et al., 2010). The size of the private sector relative to the state sector is therefore a useful indicator of local institutional quality, in that institutional quality has been a constraint on the growth of this sector.

In Chinese governance, prefectures are one administrative level below provinces, and a broader range of prefectural-level economic data is published at this level than at smaller, subordinate levels (i.e., the county level). In an effort to determine whether discontinuities in this measure of institutional quality exist at provincial borders, I construct a series of comparisons based on prefectures adjacent to each provincial border. Figure 6 represents the site of one such comparison, at the Guangdong-Hunan border. Among Hunan's 14 prefectures, Yongzhou and Chenzhou are closest to the border with Guangdong. Among Guangdong's 21 prefectures, Qingyuan and Shaoguan are closest to the border with Hunan. These prefectures are highlighted in the figure. The location of provincial capitals, Changsha and Guangzhou, are indicated. The intuition behind limiting the comparison of private sector employment to prefectures along the border is the same as that underlying spatial regression discontinuity designs generally, insofar as province-level variation is best isolated at locations as near as possible to a provincial border. By this logic, data on private sector employment at smaller administrative levels would be preferable, but prefectural data is the best available.

**Insert Figure 6 about here**

To operationalize private sector employment at the prefectural level, I divide the category, "Number of employed persons in urban private enterprises and self-employed individuals" by the category "employees", the latter referring to all employees within the prefecture. I call the resulting variable *percent private*. At each border, I then compute the average of this variable across prefectures on either side of the border[12]. Finally, a dataset consisting of dyads of border regions is created, and for each pair of regions, the difference in *percent private* is calculated. I also calculate the difference in rankings of institutional quality ascribed to each province in the pair by the World Bank. The difference in *percent private* is calculated such that its value in the purportedly low-institutional quality province is subtracted from that of the high quality province. Similarly, note that institutional quality rankings are reverse-ordered, such that the highest-rated province, Guangdong, is ranked as 30, and the lowest ranked province, Gansu, is ranked as 1. Figure 7 is a scatter plot of the association between differences in *percent private* and differences in rank. In these smaller administrative areas adjacent to provincial borders, increasing province-level differences in institutional quality are associated with increasing differences in one measure of institutional quality, the percent of the labour force employed in private enterprises. The association is significant at the .05 level.

**Insert Figure 7 about here**

Though neither of these tests of the assumption that variation in institutional quality is in fact discontinuous at provincial borders is by itself conclusive, I view the consistency of this evidence, obtained from a multiple sources using varied definitions of institutional quality, as strongly supportive of this assumption.

---

[12] Data on private sector employment is not collected among autonomous prefectures. These are prefectures where, generally speaking, a majority of the population is comprised of ethnic minority (i.e., non-Han Chinese) residents. At borders where autonomous prefectures are among the adjacent prefectures, these are omitted from the construction of the variable. At borders where all adjacent prefectures in one or both provinces are autonomous prefectures, this border pair of provinces is omitted from the analysis.

*Main results: Comparability across observables*

I begin by demonstrating both graphically and via RDD estimation the validity of the assumption of comparability of border regions with respect to ethnicity, precipitation, elevation and road density.  In terms of geography and climate, Figure 8 shows that provincial borders do not track differences in elevation, which increases as one moves westward, while Figure 9 shows that July precipitation levels vary smoothly across provincial borders, also increasing as one moves west.

**Insert Figure 8 about here**

**Insert Figure 9 about here**

Tables 2 through 5 demonstrate comparability using local linear regressions with a linear polynomial in which covariates (elevation, precipitation, distance to roads and population ) are treated as outcome variables. In the linear specification, there are no significant discontinuities in controls at borders at p = 0.05.  In the quadratic specifications (available in the online appendix), precipitation at the Anhui-Henan border and the Xinjiang-Gansu border differ, as does road density at the Hubei-Hunan border.  Though these covariates are included most specifications, results at these borders should be interpreted with greater caution.

**Insert Tables 2 through 5 about here**

*Main results: Institutional quality and income*

Main estimates derived from local linear regressions with dialect group fixed effects are reported in Table 6.  Columns (1) and (2) take *luminosity* as the outcome variable, and employ linear and quadratic polynomials, respectively.  In columns (3) and (4), the dependent variable is *luminosity* divided by cell population, again employing linear and quadratic polynomials, respectively.  In columns (5) and (6), the dependent variable is the dummy *Lit*, which takes the value of 1 when any luminosity is recorded in the cell. Linear and quadratic polynomials are employed in the models for columns (5) and (6), respectively.  In no specifications is there a significant effect of institutional quality on income. Indeed, the point estimates are of the opposite sign predicted by an institution-centric account. Regression discontinuity plots of columns (1) and (2) are presented in the upper panels of Figure 10. Cubic and quatric polynomial fits are presented in the lower panels. No discontinuity at the cutoff (border) is evident.[13]

**Insert Table 6 about here**

**Insert Figure 10 about here**

Table 7 repeats the analysis of Table 6 with control variables included. Given the above-demonstrated comparability of observed covariates across borders, these results should be interpreted as a robustness check on the main results from Table 6. For the quadratic polynomial models for luminosity and luminosity per person, I estimate a marginally statistically significant (p=.09 and p=.06, respectively) *negative* effect of institutional quality on income.

**Insert Table 7 about here**

RDD estimates at 22 borders where a putatively large discontinuity in institutional quality exists are presented in Tables 8 and 9, taking *luminosity* as the dependent variable and employing linear and quadratic polynomials, respectively. All models include covariates. At only 1 of the 22 borders in Table 8 is there a statistically significant (p = 0.05) discontinuity in *luminosity*, and this in the opposite direction as predicted by the institutionalist account. In the model employing the quadratic term in Table 9, no significant differences in *luminosity* are observed. In Tables 10 and 11, the linear and quadratic models are used in a linear probability framework to estimate the effect of institutions on *lit*, a dummy taking the value of 1 when any luminosity is recorded in a cell. All models include covariates. In no cases is there a discontinuity in the probability of illumination. This is also evident from the regression discontinuity plots in Figures 11 through 13 which again fit models of varied polynomials to identify any effects of institutional quality on *luminosity* at borders.

**Insert Tables 8 and 9 about here**

---

[13] RDD plots generated using Stata's "rdplot" command. See note 9 above for details.

**Insert Tables 10 and 11 about here**

**Insert Figures 11 through 13 about here**

### V.    Discussion

This paper extends the spatial regression discontinuity design approach to the study of subnational institutions in China.  Four broad conclusions emerge: (*i*) institutions of governance are indeed province level in that there is evidence of their discontinuity at provincial borders; (*ii*) spatial RDDs find no discontinuities at provincial borders for a range of observed covariates, suggesting the validity of the identifying assumptions of the RDD; (*iii*) estimates of the within-dialect group impact of high quality institutions on income, as proxied by three operationalizations of nighttime lights emissions, find no positive effects distinguishable from zero; (*iv*) these results are reproduced at 22 individual borders at which gaps in institutional quality as assessed by the World Bank are putatively large.  The unrestrictive identification assumptions of regression discontinuity designs, the data driven approach to bandwidth selection employed here, and the consistency of this finding across multiple specifications, argue for the robustness of these results.

Though the measure of institutional quality employed here is a de facto assessment of the ease of enforcing contracts, it should be emphasized that it is a measure of the ability or willingness of government to establish secure rights of control and transfer over property.  This is in keeping with the dominant view of the institutionalist literature that the key precondition of effective property rights is the rise of a state that is strong enough in its monopoly on force and law to secure these rights, while at the same time being credibly constrained from expropriating private property or limiting access to secure property rights to elite allies.  Our results need not imply that contracting is unimportant, nor that private enforcement of contracts is impossible, and a sizable literature has demonstrated the relevance of trust and social capital for economic growth (e.g., Algan and Cahuc, 2010).  In the Chinese context, the role of *guanxi*, a web of (often kinship) relationships based on mutual moral obligations extending through time, in securing property against government predation has been documented (Zhang and Zhao, 2014).  The ability to invest in these informal structures to better secure property rights implies that in the absence of high quality government-backed institutions, economic agents can nevertheless reduce transaction costs to a degree. If these informal structures have been an important basis for contracting in reform-era China, this speaks to the importance of trust networks and familial relationships, rather than government-backed institutions, and trust networks that are not limited by provincial boundaries. Thus, even if trust networks substitute for formal structures in supporting economic exchange in China, this argues against the importance of institutions as commonly understood by economists.

The paper contributes to a growing literature which has found either an absence of support for institution-centric accounts of comparative development (e.g., Michalopoulos and Papaioannou, 2014), or support for counter-narratives emphasizing the importance human capital, geography and culture (e.g., Galor and Ozak, 2016; Putterman and Weil, 2010; Spolaore and Wacziarg (2009, 2013). This suggests, in the Chinese context and beyond, that the geographic, cultural and human capital roots of comparative development bear continued study.

These results do not imply that no change in the quality of institutions, of whatever magnitude, can have a substantial effect on incomes. The coincidence of the timing of China's rise with the elimination of Maoist era agriculture policies, restrictions on labour mobility and commercial activity generally, is unlikely to be coincidental. However, these results suggest that China's regional inequality is not the result of different development models pursued by China's provinces.

**References**

Acemoglu, D., Johnson, S., & Robinson, J. A. (2005). Institutions as a fundamental cause of long-run growth. *Handbook of economic growth*, *1*, 385-472.

Acemoglu, D., Johnson, S., & Robinson, J. A. (2002). Reversal of fortune: Geography and institutions in the making of the modern world income distribution. *The Quarterly journal of economics*, *117*(4), 1231-1294.

Acemoglu, D., & Robinson, J.A. (2001). The Colonial Origins of Comparative Development: An Empirical Investigation. *The American Economic Review*, *91*(5), 1369-1401.

Addison, D. M., & Stewart, B. (2015). Nighttime lights revisited: the use of nighttime lights data as a proxy for economic variables. *World Bank Policy Research Working Paper*, (7496).

Algan, Y., & Cahuc, P. (2010). Inherited Trust and Growth. *American Economic Review*, *100*(5), 2060-92.

Almås, I., Johnsen, Å. A., & Kotsadam, A. (2014). *Poverty in China seen from outer space* (No. 11/2014). Memorandum, Department of Economics, University of Oslo.

Athey, S., & Imbens, G. (2016). The State of Applied Econometrics: Causality and Policy Evaluation. *Journal of Economic Perspectives*, 31(2), 3-32.

Bai, Y., & Kung, J. K. S. (2015). Does Genetic Distance have a Barrier Effect on Technology Diffusion? Evidence from Historical China.

Bai, Y., & Kung, J. K. S. (2011). Genetic distance and income difference: Evidence from changes in China's cross-strait relations. *Economics Letters*, *110*(3), 255-258.

Balding, C. (2013). How Badly Flawed is Chinese Economic Data? The Opening Bid is $1 Trillion. *The Opening Bid is $1 Trillion (August 14, 2013)*.

Bao, S., Chang, G. H., Sachs, J. D., & Woo, W. T. (2002). Geographic factors and China's regional development under market reforms, 1978–1998. *China Economic Review*, *13*(1), 89-111.

Berman, L., Zhang, W. (2017). V6 Ming dynasty courier routes and stations. Harvard Dataverse. doi:10.7910/DVN/SB8ZTM

Birney, M. (2014). Decentralization and veiled corruption under China's "rule of mandates". *World Development*, *53*, 55-67.

Calonico, S., Cattaneo, M. D., & Titiunik, R. (2014). Robust data-driven inference in the regression-discontinuity design. *Stata Journal*, *14*(4), 909-946.

Calonico, S., Cattaneo, M. D., Farrell, M. H., & Titiunik, R. (2016). rdrobust: Software for regression discontinuity designs. Forthcoming in *The Stata Journal.*

Calonico, S., Cattaneo, M. D., Farrell, M. H., & Titiunik, R. (2018). Regression discontinuity designs using covariates. *Review of Economics and Statistics*, (0).

Center for International Earth Science Information Network. (2013). Global roads open access database, version 1 (gROADSv1). Palisades, NY: NASA Socioeconomic Data and Applications Center (SEDAC). http://dx.doi.org/10.7927/H4VD6WCT.

Center for International Earth Science Information Network. (2016). Gridded Population of the World, Version 4 (GPWv4): Population count adjusted to Match 2015 revision of UN WPP country totals. Palisades, NY: NASA Socioeconomic Data and Applications Center (SEDAC). http://dx.doi.org/10.7927/H4SF2T42.

Chen, C. J., Li, Z., Su, X., & Sun, Z. (2011). Rent-seeking incentives, corporate political connections, and the control structure of private firms: Chinese evidence. *Journal of Corporate Finance*, *17*(2), 229-243.

Chen, S., & Lan, X. (2017). There will be killing: Collectivization and death of draft animals. *American Economic Journal: Applied Economics*, *9*(4), 58-77.

Chen, X., & Nordhaus, W. D. (2011). Using luminosity data as a proxy for economic statistics. *Proceedings of the National Academy of Sciences*, *108*(21), 8589-8594.

Choi, E. K. (2009). The politics of fee extraction from private enterprises, 1996-2003. *The China Journal*, (62), 79-102.

Chow, G. (2006). Are Chinese official statistics reliable?. *CESifo Economic Studies*, *52*(2), 396-414.

Clark, G. (2007). A Review of Avner Greif's Institutions and the Path to the Modern Economy: Lessons from Medieval Trade. *Journal of Economic Literature*, *45*(3), 725-741.

Comin, D., Easterly, W., & Gong, E. (2010). Was the Wealth of Nations determined in 1000 BC? *American Economic Journal: Macroeconomics*, *2*(3), 65-97.

Cook, T. D., Shadish, W. R., & Wong, V. C. (2008). Three conditions under which experiments and observational studies produce comparable causal estimates: New findings from within-study comparisons. *Journal of policy analysis and management*, *27*(4), 724-750.

Cull, R., & Xu, L. C. (2005). Institutions, ownership, and finance: the determinants of profit reinvestment among Chinese firms. *Journal of Financial Economics*, *77*(1), 117-146.

Dell, M. (2010). The persistent effects of Peru's mining mita. *Econometrica*, *78*(6), 1863-1903.

Dollar, D., Wang, S., Xu, L., & Shi, A. (2004). Improving city competitiveness through the investment climate: Ranking 23 Chinese cities.

Du, J., Lu, Y., & Tao, Z. (2008). Economic institutions and FDI location choice: Evidence from US multinationals in China. *Journal of comparative Economics*, *36*(3), 412-429.

Edin, M. (2003). State capacity and local agent control in China: CCP cadre management from a township perspective. *The China Quarterly*, *173*, 35-52.

Fairbank Center for Chinese Studies and the Institute for Chinese Historical Geography at Fudan University. (2012). CHGIS, Version: 5.

Fick, S. E., & Hijmans, R. J. (2017). WorldClim 2: new 1-km spatial resolution climate surfaces for global land areas. *International Journal of Climatology*, *37*(12), 4302-4315.

Fitzgerald, J. (2002). The province in history. *Rethinking China's provinces*, 11-40.

Fuller, G. H. (2007). Economic Warlords: How De Facto Federalism Inhibits China's Compliance with International Trade Law and Jeopardizes Global Environmental Initiatives. *Tenn. L. Rev.*, *75*, 545.

Galor, O., & Özak, Ö. (2015). Land productivity and economic development: Caloric suitability vs. agricultural suitability.

Galor, O., & Özak, Ö. (2016). The agricultural origins of time preference. *The American Economic Review*, *106*(10), 3064-3103.

Glaeser, E. L., La Porta, R., Lopez-de-Silanes, F., & Shleifer, A. (2004). Do institutions cause growth?. *Journal of economic Growth*, *9*(3), 271-303.

Hall, R. E., & Jones, C. I. (1999). Why do some countries produce so much more output per worker than others?. *The quarterly journal of economics*, *114*(1), 83-116.

Hallward-Driemeier, M., Wallsten, S., & Xu, L. C. (2006). Ownership, investment climate and firm performance. *Economics of Transition*, *14*(4), 629-647.

Henderson, J. Vernon, Adam Storeygard, and David N. Weil. (2012). "Measuring Economic Growth from Outer Space." *American Economic Review*, *102*(2): 994-1028.

Huang, Y. (2008). *Capitalism with Chinese characteristics: Entrepreneurship and the state*. Cambridge University Press.

Imbens, G. W., & Lemieux, T. (2008). Regression discontinuity designs: A guide to practice. *Journal of econometrics*, *142*(2), 615-635.

Jerven, M. (2013). *Poor numbers: how we are misled by African development statistics and what to do about it*. Cornell University Press.

Keele, L. J., & Titiunik, R. (2015). Geographic boundaries as regression discontinuities. *Political Analysis*, *23*(1), 127-155.

Keele, L., & Titiunik, R. (2016). Natural experiments based on geography. *Political Science Research and Methods*, *4*(1), 65-95.

Keola, S., Andersson, M., & Hall, O. (2015). Monitoring economic development from space: using nighttime light and land cover data to measure economic growth. *World Development*, *66*, 322-334.

Klemens, B., Coppola, A., & Shron, M. (2015). Estimating local poverty measures using satellite images: a pilot application to Central America.*World Bank Policy Research Working Paper*, (7329).

Koch-Weser, I. N. (2013). The reliability of China's economic data: An analysis of national output. *US-China Economic and Security Review Commission Staff Research Project*.

Knack, S., & Keefer, P. (1995). Institutions and economic performance: cross-country tests using alternative institutional measures. *Economics & Politics*, *7*(3), 207-227.

Lardy, N. (2006). China: Toward a consumption-driven growth path.

Lavely, W. (2001). Coding Scheme for the Language Atlas of China. *University of Washington Center for Studies in Demography and Ecology Working Papers*, 01-7.

Lavely, W., & Berman, L. (2012). Language Atlas of China. Harvard Dataverse, V1. https://doi.org/10.7910/DVN/QPUONU .

Li, C., & Gibson, J. (2013). Rising regional inequality in China: Fact or artifact?. *World Development*, *47*, 16-29.

Lin, C., Lin, P., & Song, F. (2010). Property rights protection and corporate R&D: Evidence from China. *Journal of Development Economics*, *93*(1), 49-62.

Lin, J. Y., & Liu, P. (2006). *Development strategies and regional income disparities in China* (No. 2006/129). Research Paper, UNU-WIDER, United Nations University (UNU).

Lu, Y., Png, I. P., & Tao, Z. (2013). Do institutions not matter in China? Evidence from manufacturing enterprises. *Journal of Comparative Economics*, *41*(1), 74-90.

Mako, W. (2006). *China: governance, investment climate, and harmonious society: competitiveness*

*enhancements for 120 cities in China*. World Bank.

Michalopoulos, S., & Papaioannou, E. (2014). National Institutions and Subnational Development in Africa. *The Quarterly Journal of Economics*,*129* (1), 151-213.

Michalopoulos, S., & Papaioannou, E. (2017). *Spatial Patterns of Development: A Meso Approach* (No. w24088). National Bureau of Economic Research.

Min, B., & Gaba, K. M. (2014). Tracking Electrification in Vietnam Using Nighttime Lights. *Remote Sensing*, *6*(10), 9511-9529.

Montinola, G., Qian, Y., & Weingast, B. R. (1995). Federalism, Chinese style: the political basis for economic success in China. *World politics*, *48*(01), 50-81.

Noor, A. M., Alegana, V. A., Gething, P. W., Tatem, A. J., & Snow, R. W. (2008). Using remotely sensed night-time light as a proxy for poverty in Africa. *Population Health Metrics*, *6*(1), 1-13.

North, D. C. (1990). *Institutions, institutional change and economic performance*. Cambridge university press.

North, D. C. (1981). *Structure and change in economic history*. Norton.

Pearson, M. M. (2007). Governing the Chinese economy: Regulatory reform in the service of the state. *Public Administration Review*, *67*(4), 718-730.

Pinkovskiy, M. L. (2017). Growth discontinuities at borders. *Journal of Economic Growth*, *22*(2), 145-192.

Poncet, S., & Barthélemy, J. (2008). China as an integrated area?. *Journal of Economic Integration*, 896-926.

Poncet, S., Steingress, W., & Vandenbussche, H. (2010). Financial constraints in China: firm-level evidence. *China Economic Review*, *21*(3), 411-422.

Putterman, L., & Weil, D. N. (2010). Post-1500 population flows and the long-run determinants of economic growth and inequality. *The Quarterly journal of economics*, *125*(4), 1627-1682.

Qian, Y., & Weingast, B. R. (1997). Federalism as a commitment to perserving market incentives. *The Journal of Economic Perspectives*, *11*(4), 83-92.

Ravallion, M. (2003). Measuring aggregate welfare in developing countries: How well do national accounts and surveys agree? *Review of Economics and Statistics*, *85*(3), 645-652.

Sachs, J. D. (2003). *Institutions don't rule: direct effects of geography on per capita income* (No. w9490). National Bureau of Economic Research.

Sachs, J. D. (2012). Government, Geography, and Growth: The True Drivers of Economic Development. *Foreign Affairs*, *91*(5), 142-150.

Spolaore, E., & Wacziarg, R. (2009). The diffusion of development. *The Quarterly Journal of Economics*, *124*(2), 469-529.

Spolaore, E., & Wacziarg, R. (2013). How deep are the roots of economic development?. *Journal of Economic Literature*, *51*(2), 325-369.

Tabellini, G. (2010). Culture and institutions: economic development in the regions of Europe. *Journal of the European Economic Association*, *8*(4), 677-716.

Wei, Y. D. (2007). Regional development in China: Transitional institutions, embedded globalization, and hybrid economies. *Eurasia Geography and Economics*, *48*(1), 16-36.

Weidmann, N. B., Rød, J. K., & Cederman, L. E. (2010). Representing ethnic groups in space: A new dataset. *Journal of Peace Research*.

Weingast, B. R. (1995). The economic role of political institutions: Market-preserving federalism and economic development. *Journal of Law, Economics, & Organization*, 1-31.

Wong, C. (2007). Fiscal management for a harmonious society: assessing the central government's capacity to implement national policies. *British Inter-university China Centre Working Paper*, *4*.

World Bank. 2008. *Doing Business in China 2008 (English)*. Washington DC : World Bank. http://documents.worldbank.org/curated/en/783261468020349761/Doing-Business-in-China-2008

*World development indicators. Washington, D.C. :The World Bank.*

Xianping, R. (2004). Research on China's Small and Medium-Sized Enterprises' Cluster Development Model. *Chinese economy*, *37*(5), 7-18.

Xu, C. (2011). The fundamental institutions of China's reforms and development. *Journal of Economic Literature*, *49*(4), 1076-1151.

Xu, C., & Zhang, X. (2009). *The evolution of Chinese entrepreneurial firms: Township-village enterprises revisited* (Vol. 854). Intl Food Policy Res Inst.

Yao, S., & Zhang, Z. (2001). On regional inequality and diverging clubs: a case study of contemporary China. *Journal of Comparative Economics*, *29*(3), 466-484.

Ye, X., & Wei, Y. D. (2005). Geospatial analysis of regional development in China: the case of Zhejiang Province and the Wenzhou model. *Eurasia Geography and Economics*, *46*(6), 445-464.

Young, S. (1989). Policy, practice and the private sector in China. *The Australian Journal of Chinese Affairs*, (21), 57-80.

Zhang, T., & Zhao, X. (2014). Do kinship networks strengthen private property? Evidence from rural China. *Journal of Empirical Legal Studies*, *11*(3), 505-540.

Figure 1.a) Traditional ethnic homelands, China

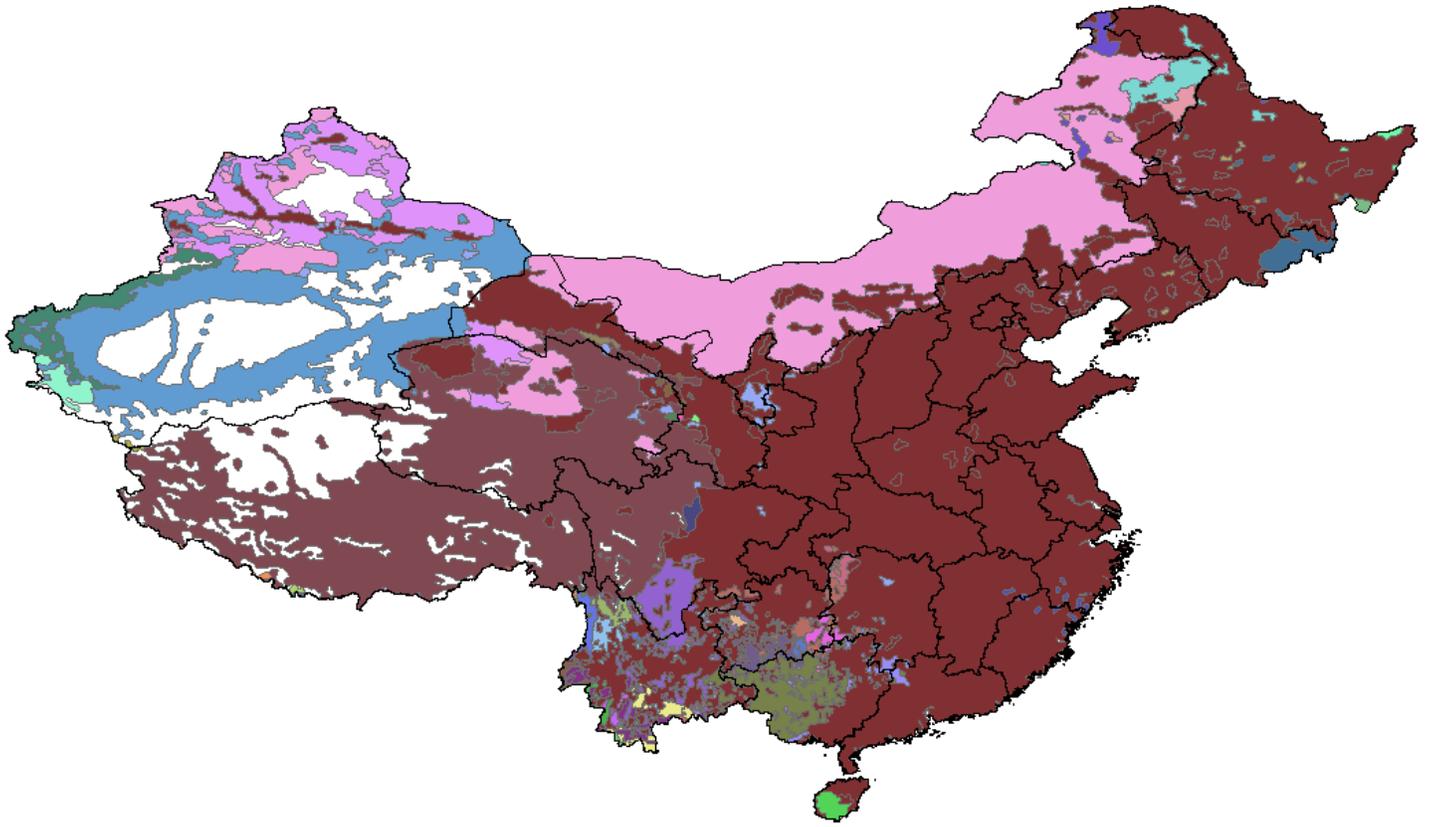

Figure 1.b) Traditional homelands of linguistic groups, China

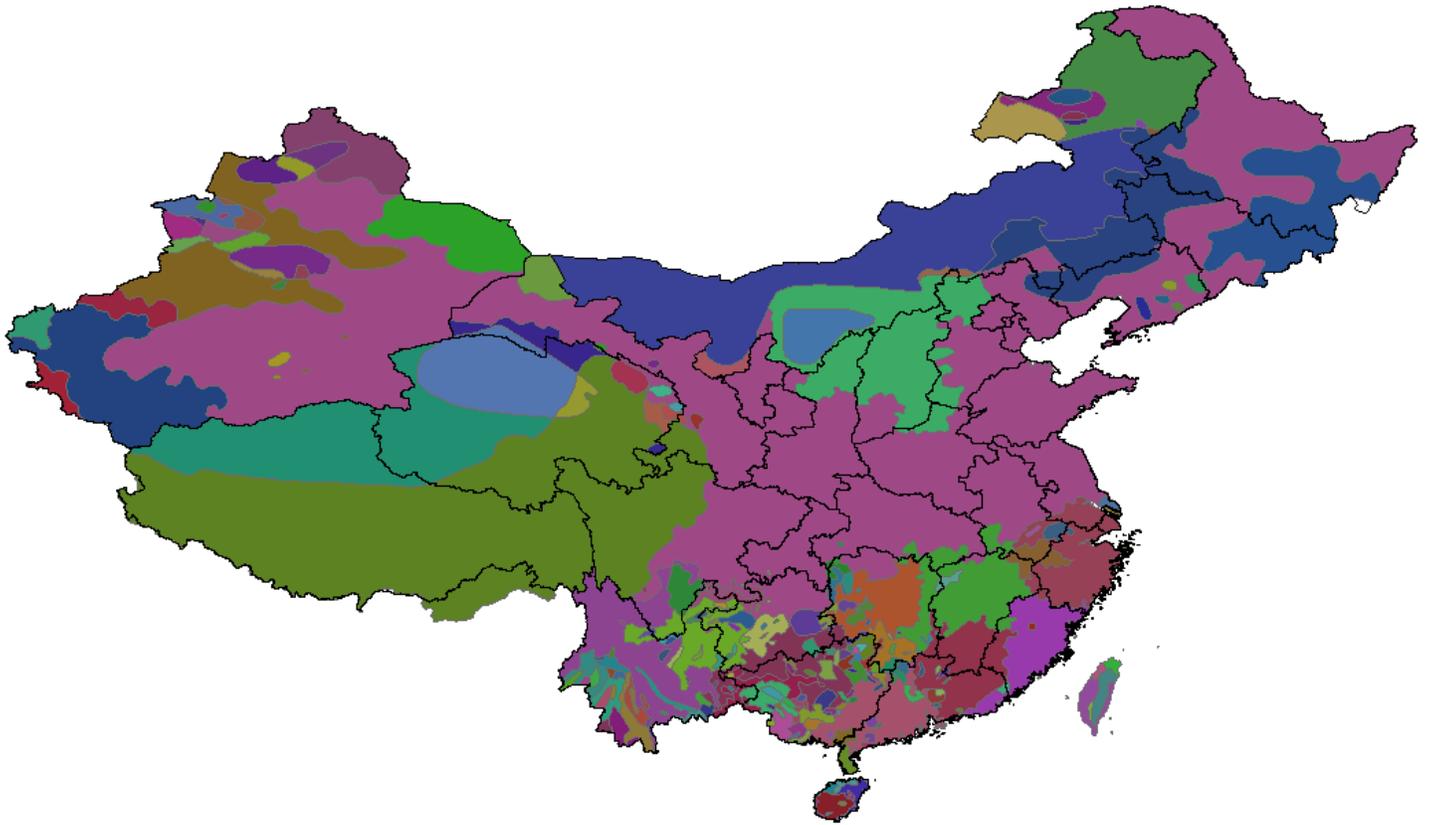

Source: Lavely and Berman (2012).

Figure 2. Within dialect group variation in institutional quality (Sichuan-Gansu border)

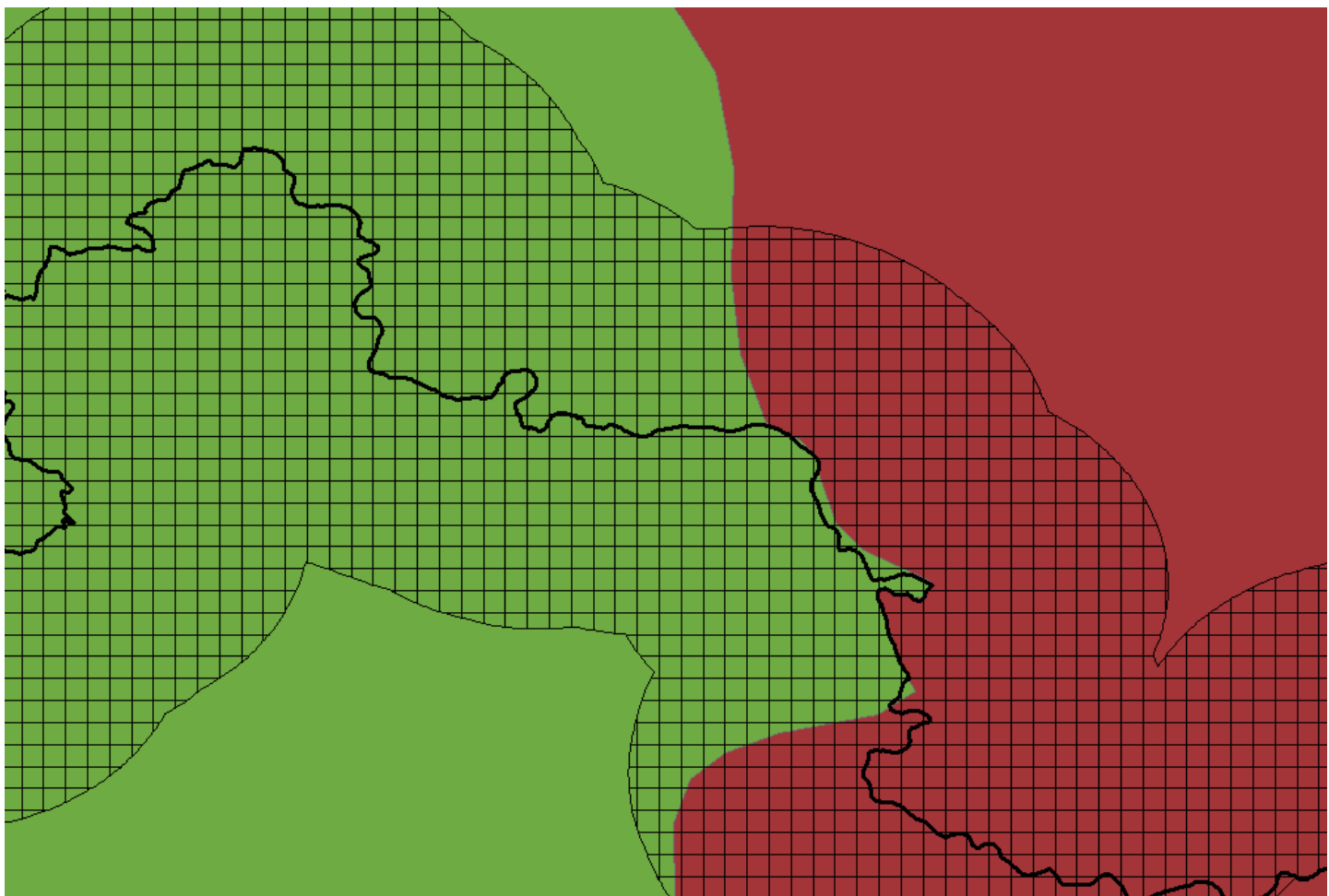

Notes: Bold line represents border between Gansu (north of border) and Sichuan (south of border). "Fishnet" indicates extent of buffer zone used to construct observations along border. Red indicates traditional homeland of speakers of dialects within the Mandarin supergroup, and greed the traditional homeland of speakers of Tibeto-Burman languages. See Lavely and Berman's (2012) Language Atlas in China.

Figure 3. Scatter plot, contract enforcement rank and per capita GDP across Chinese provinces

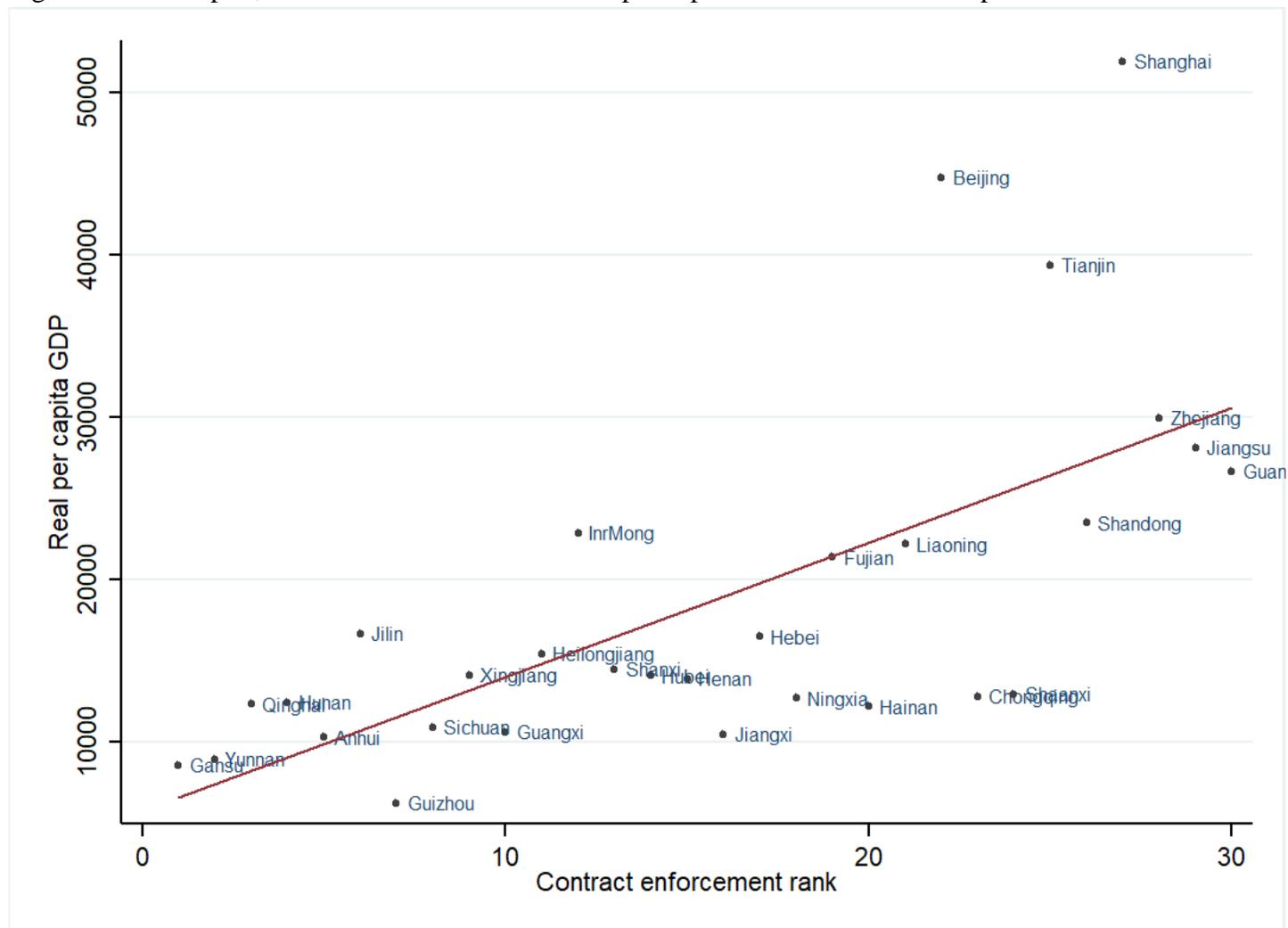

*Notes*: Contract enforcement rank is via World Bank. Per capita GDP data is from China's National Bureau of Statistics.

Figure 4a.  Luminosity within cells in buffer zones surrounding border

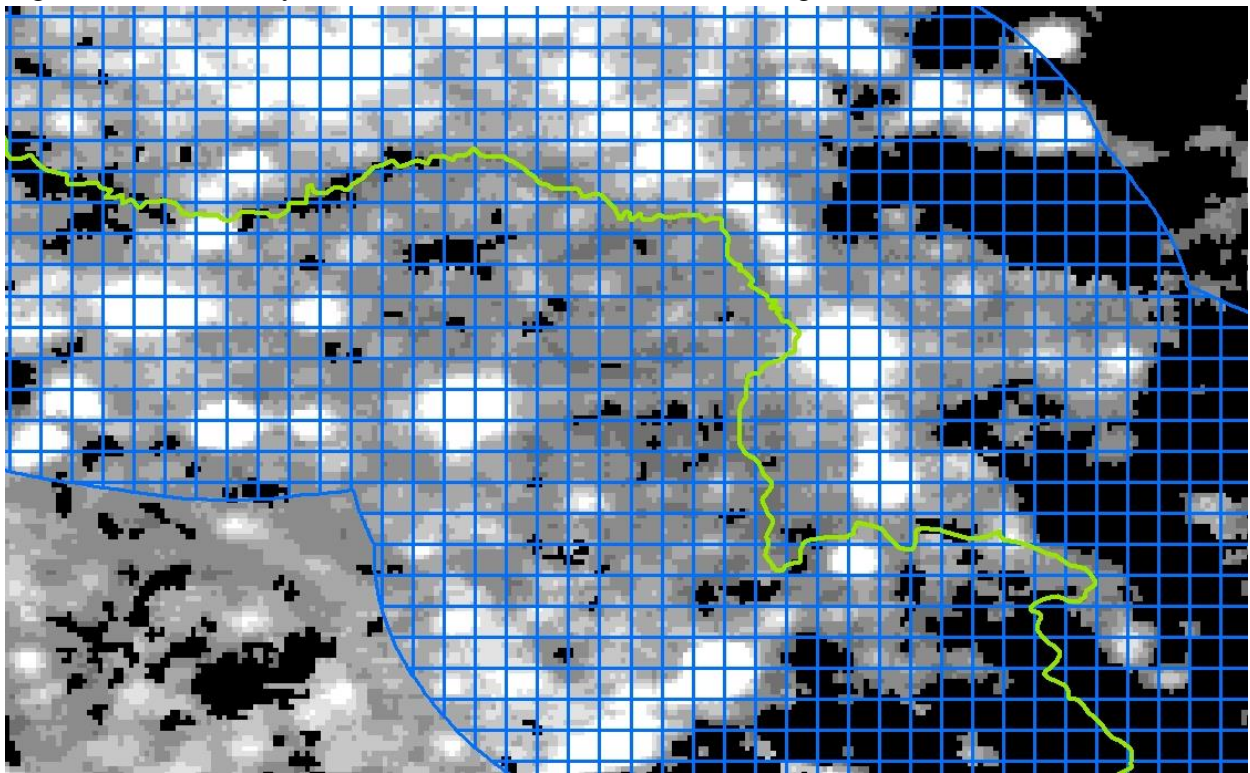

Notes: Bright green line indicates location of border between Heilongjiang and Jilin provinces.  Blue grid is a "fishnet" created in ArcGIS which defines the unit of analysis, the cell.  Extent of grid indicates 50km "buffer zone" around the border.  Cells to the north of the border are in Heilongjiang, and those to the south in Jilin.  Background colour indicates pixel-level luminosity, lighter shades reflecting higher light emissions.

Figure 4b. Luminosity within cells in buffer zones surrounding border (high zoom)

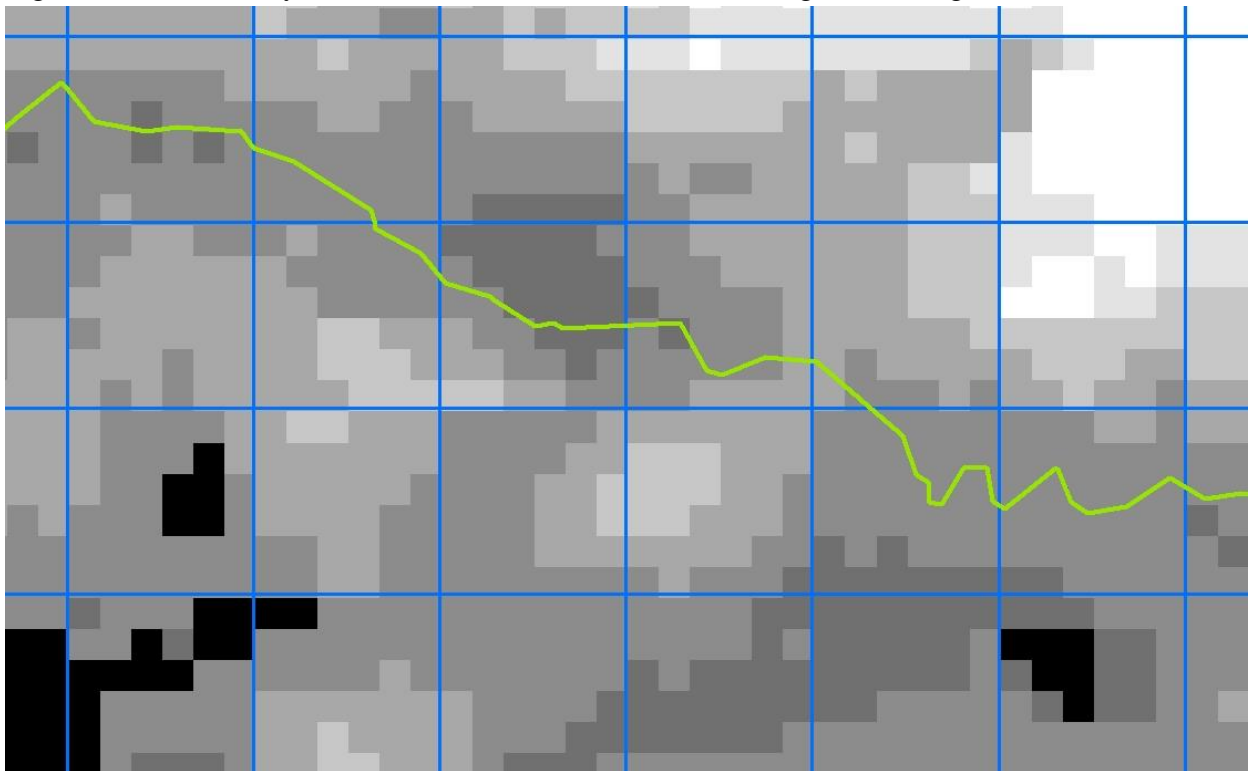

Notes: As in Figure 2a, scale 1: 125,000.

Figure 5. Within and across province differences in mean survey responses

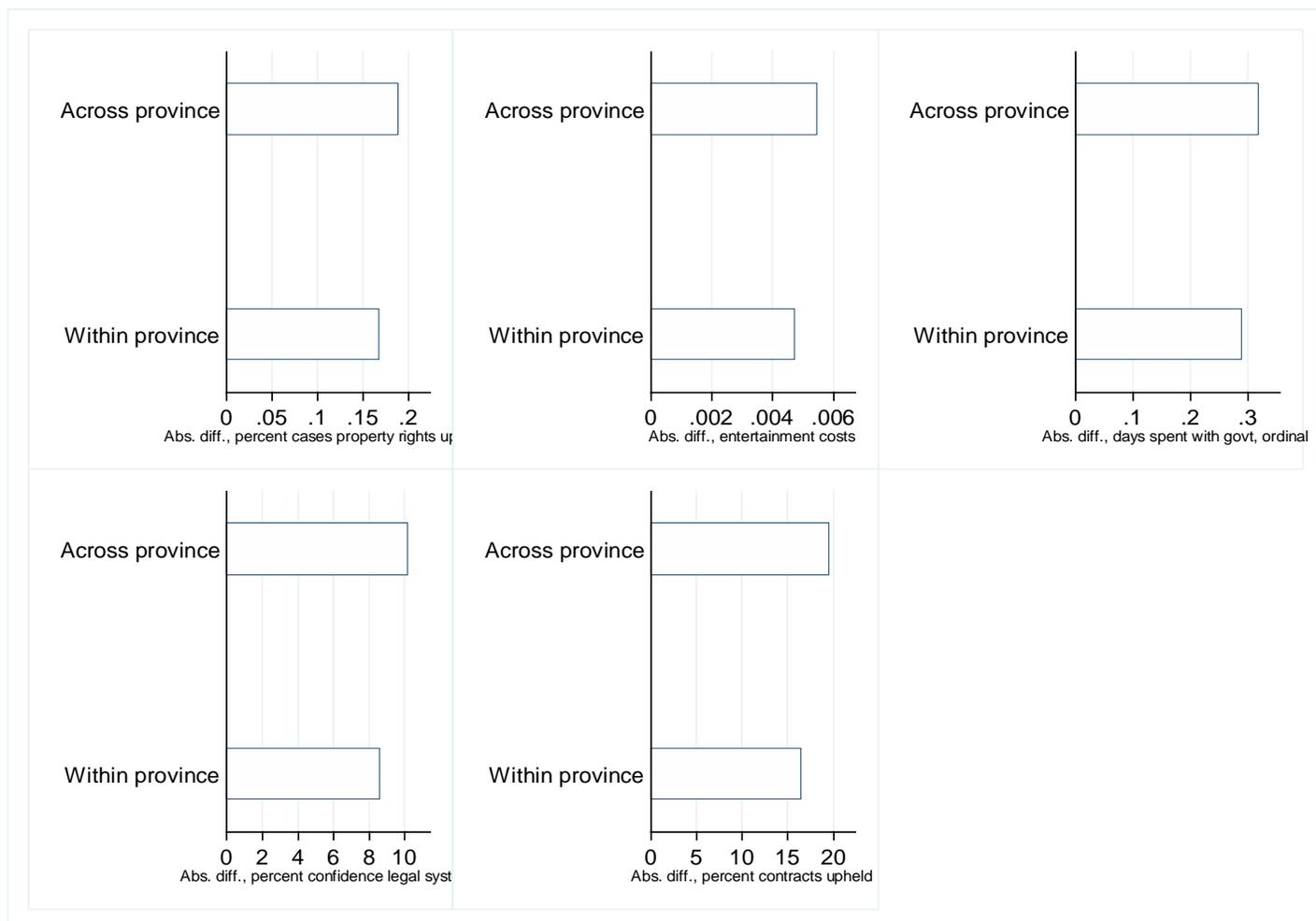

Figure 6. Setting for prefecture-level comparison of private enterprise employment

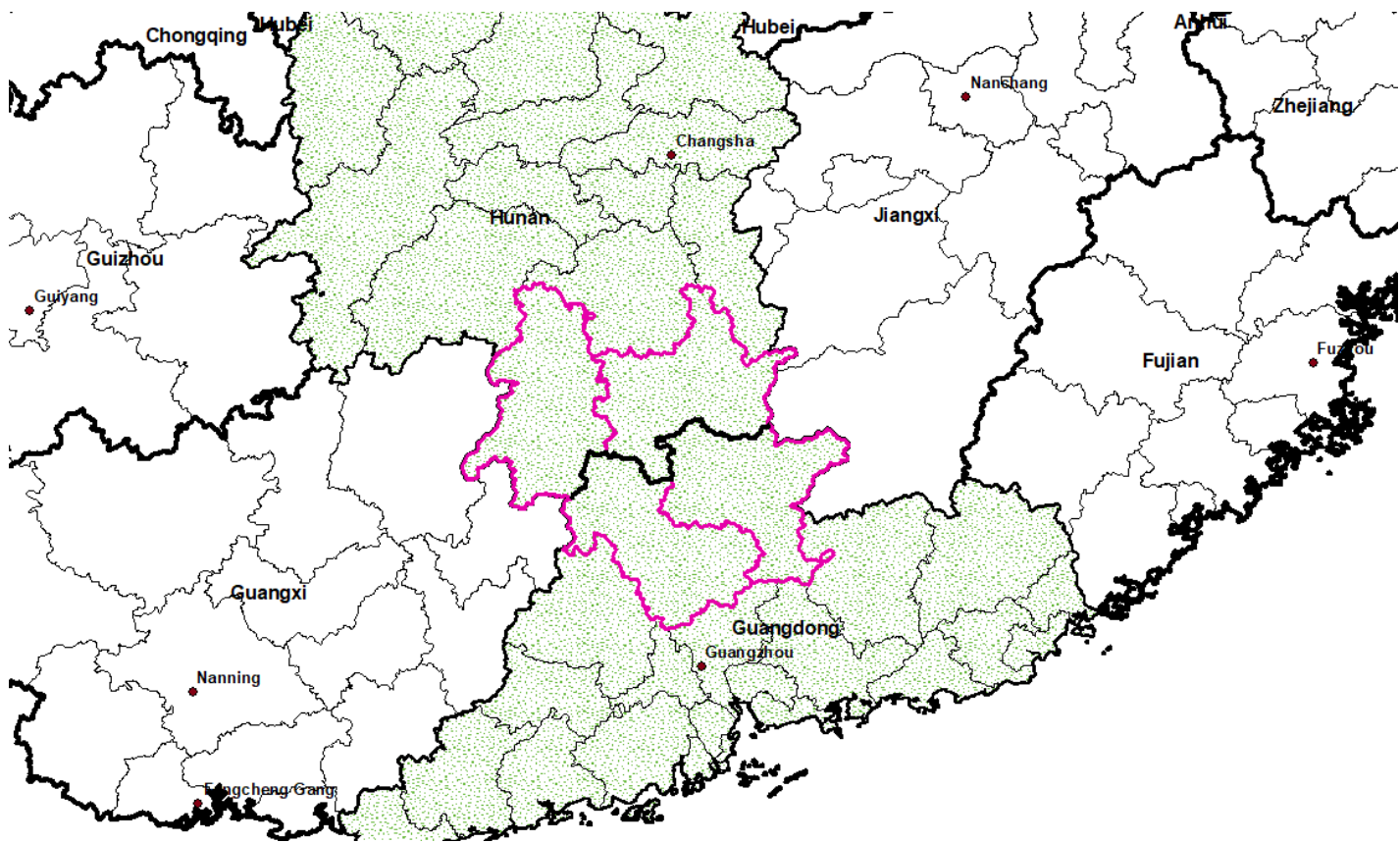

Figure 7. Scatter plot of difference in percent private sector employment and difference in institutional rank, bordering prefectures

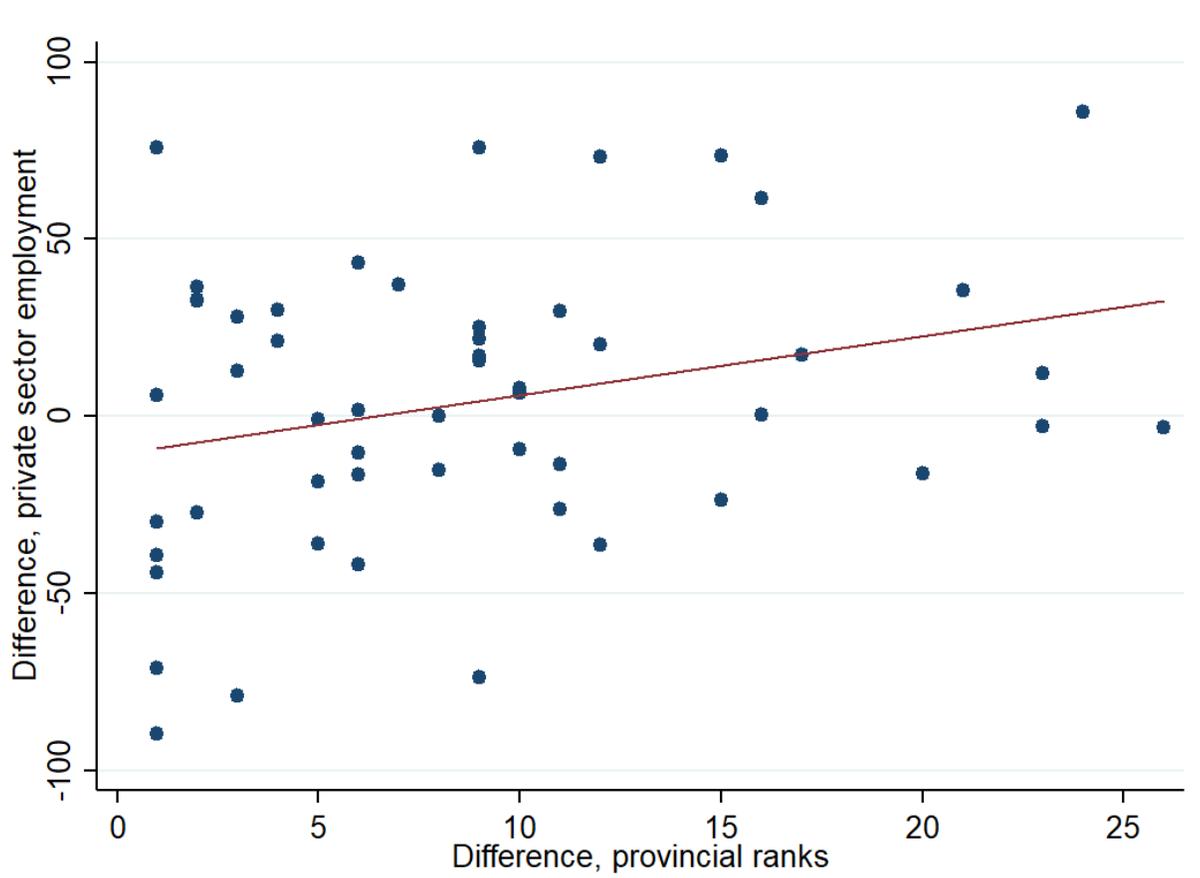

Figure 8. Elevation across Chinese provincial borders.

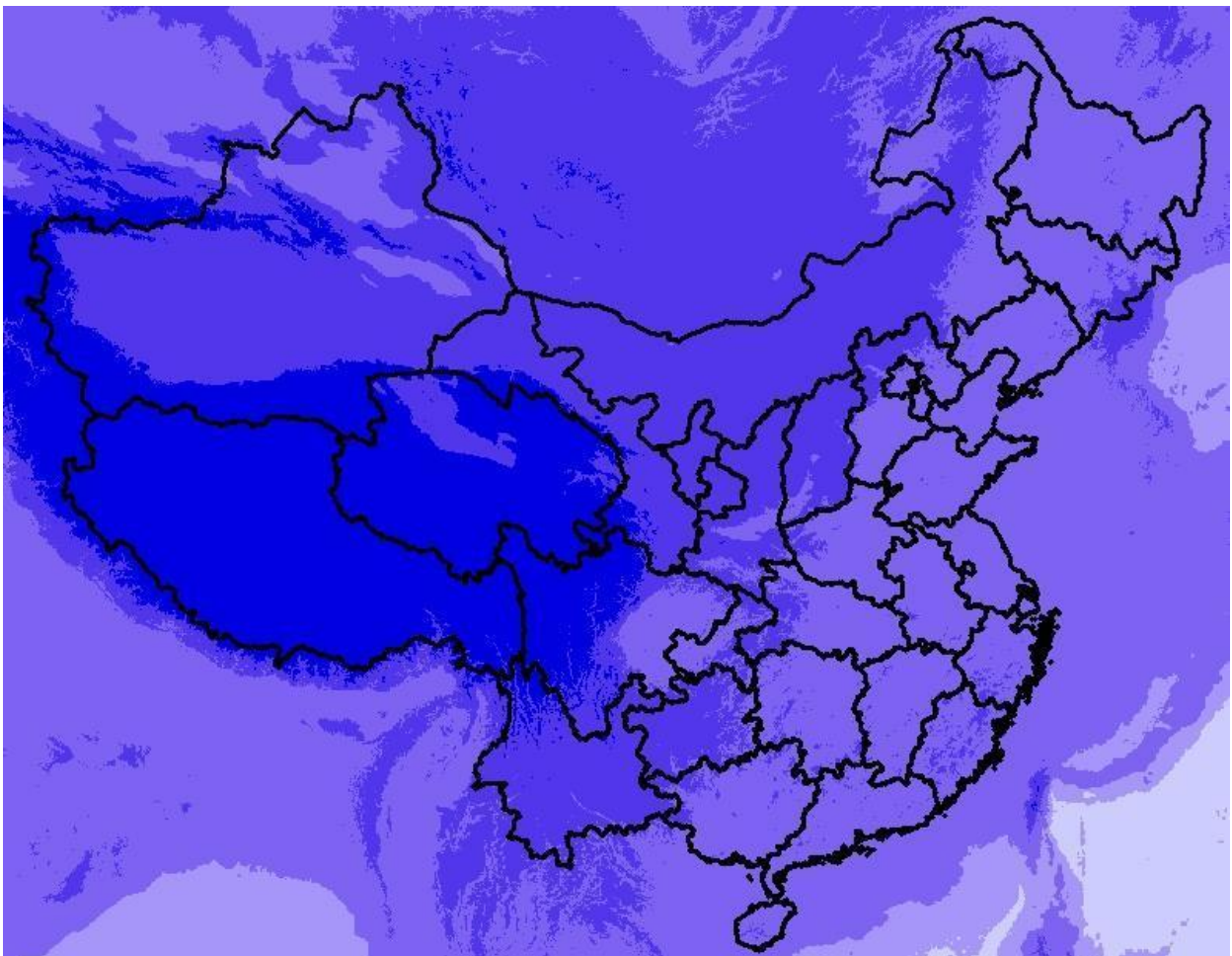

*Notes*: Data are from Harvard's China Historical Geographic Information System. Data available via
https://www.fas.harvard.edu/~chgis/data/chgis/v5/. Darker shades indicate higher elevation. Processed using ArcGIS.

Figure 9.  July precipitation across Chinese provincial borders

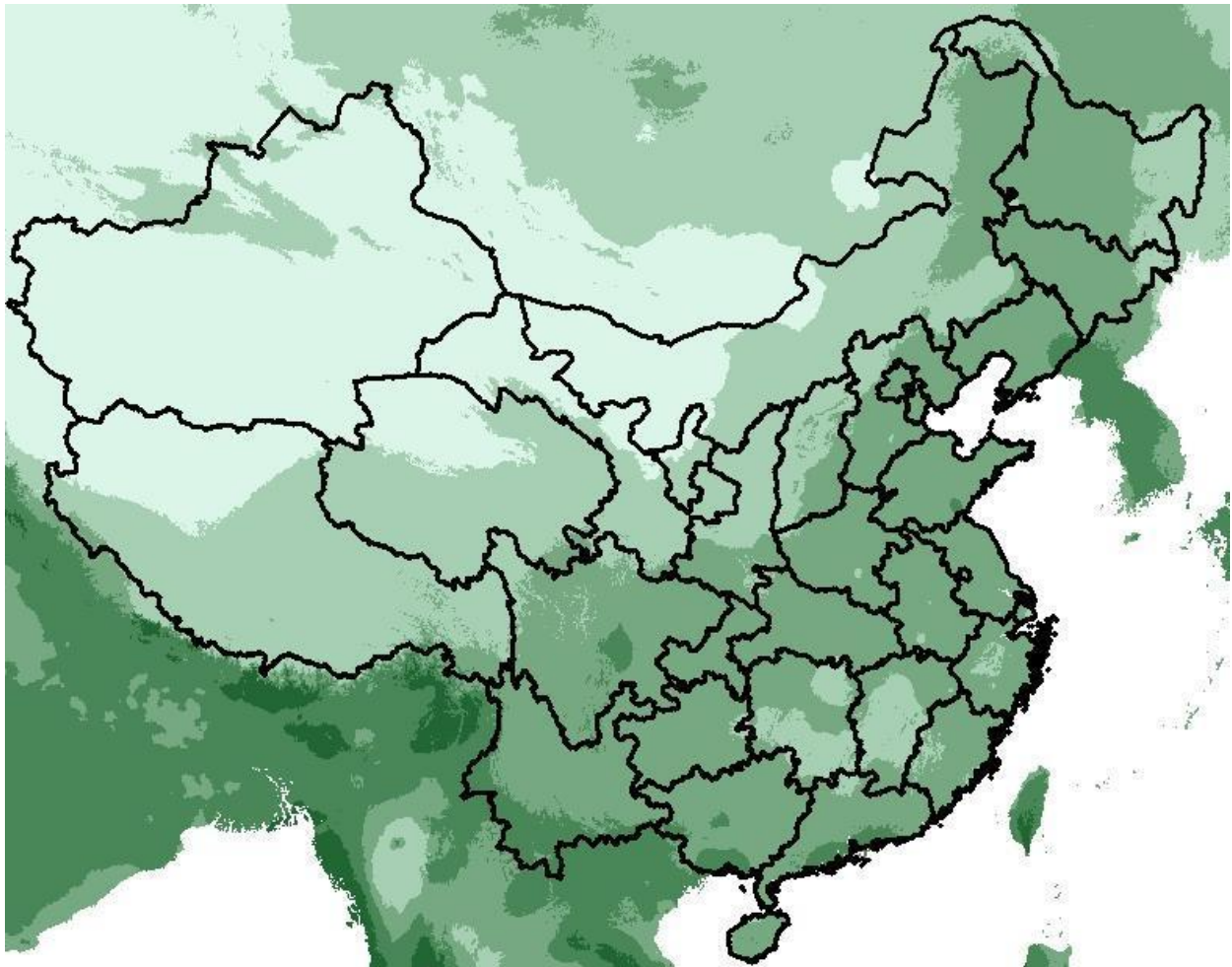

*Notes*: Data are from WorldClim – Global Climate Data.  See http://www.worldclim.org.  Darker shades indicate higher July precipitation levels.  Processed using ArcGIS.

Figure 10. Regression discontinuity plots, dialect group fixed effect estimates.

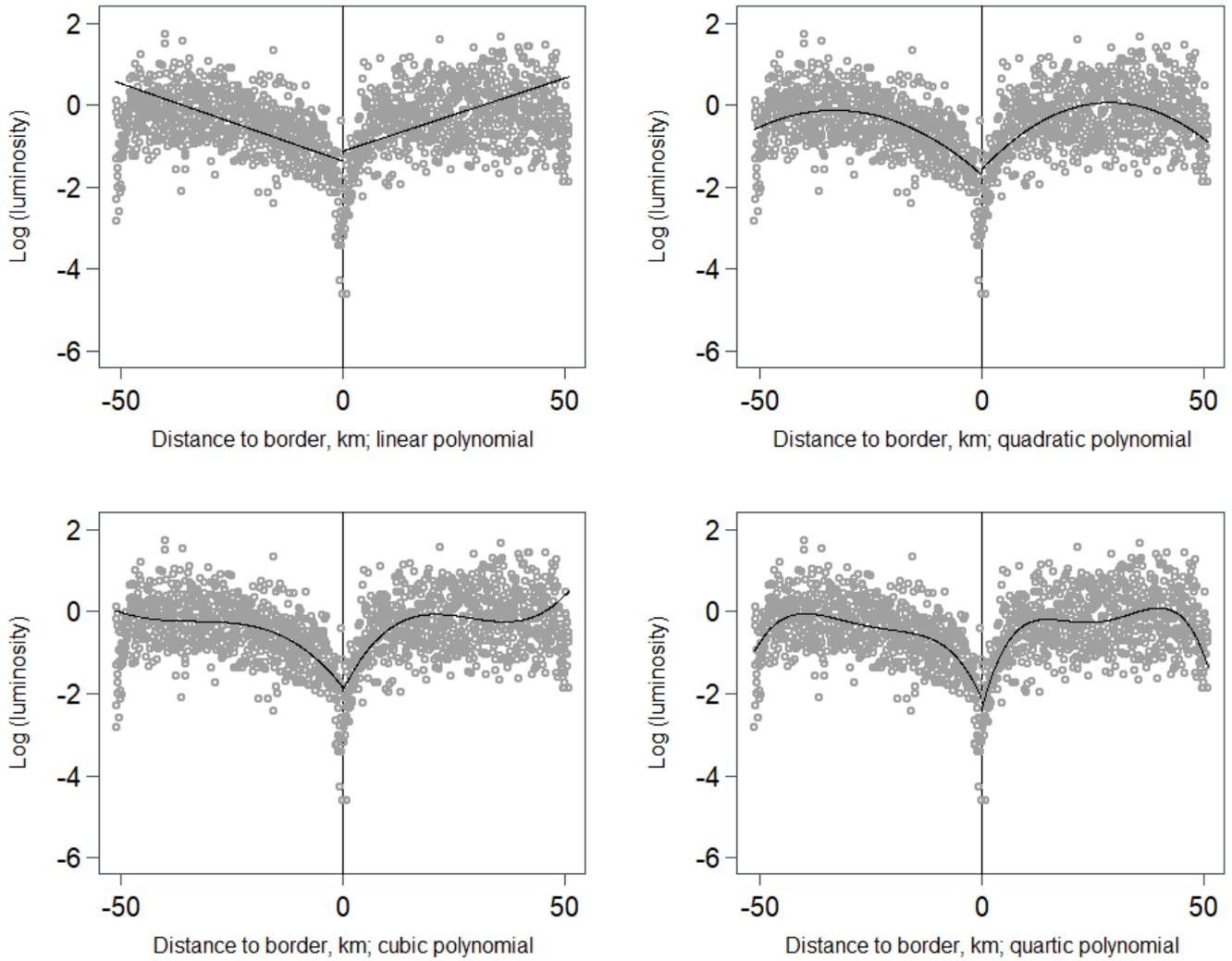

Figure 11. Regression discontinuity plots, Gansu-Ningxia border.

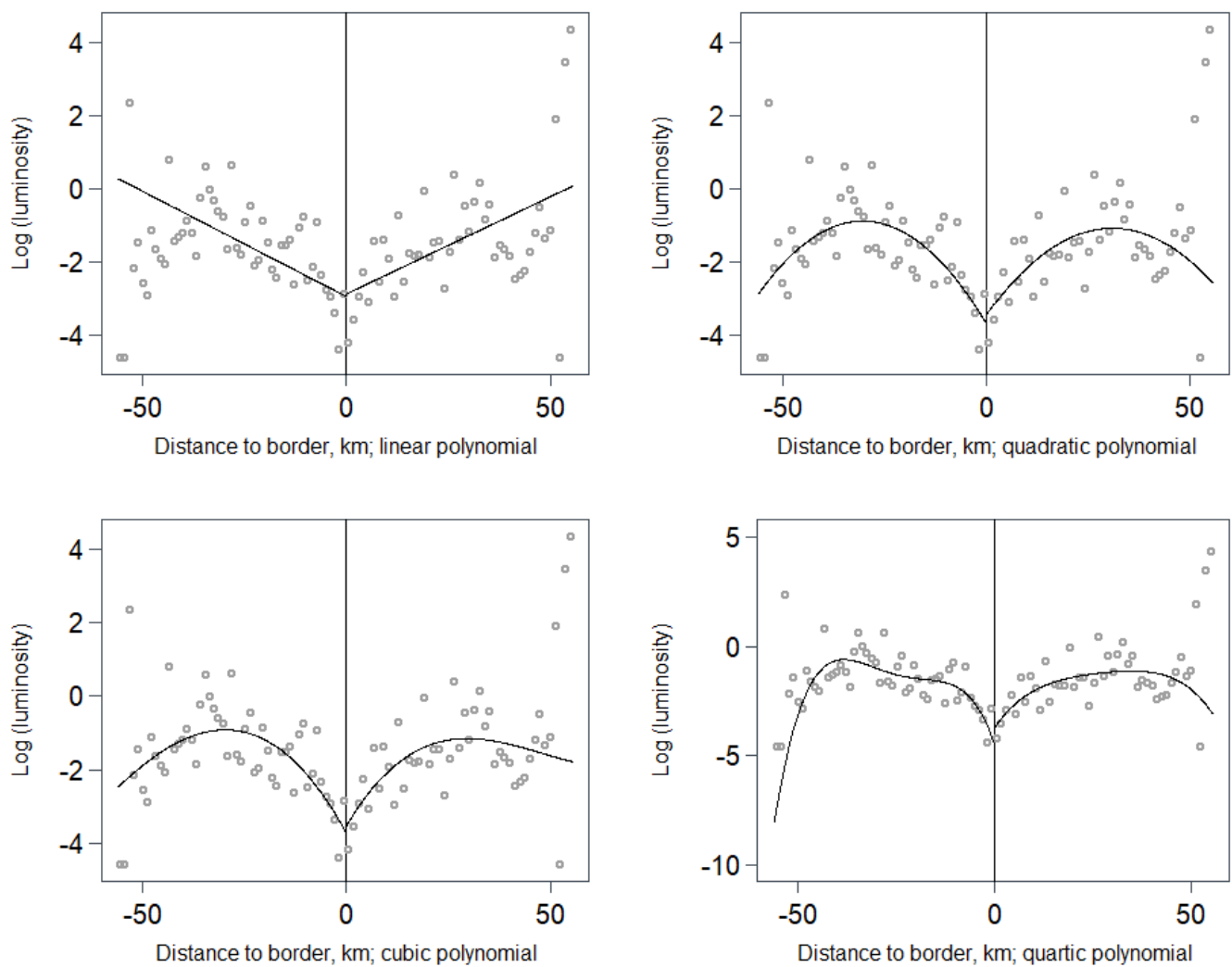

Figure 12. Regression discontinuity plots, Anhui-Jiangsu border.

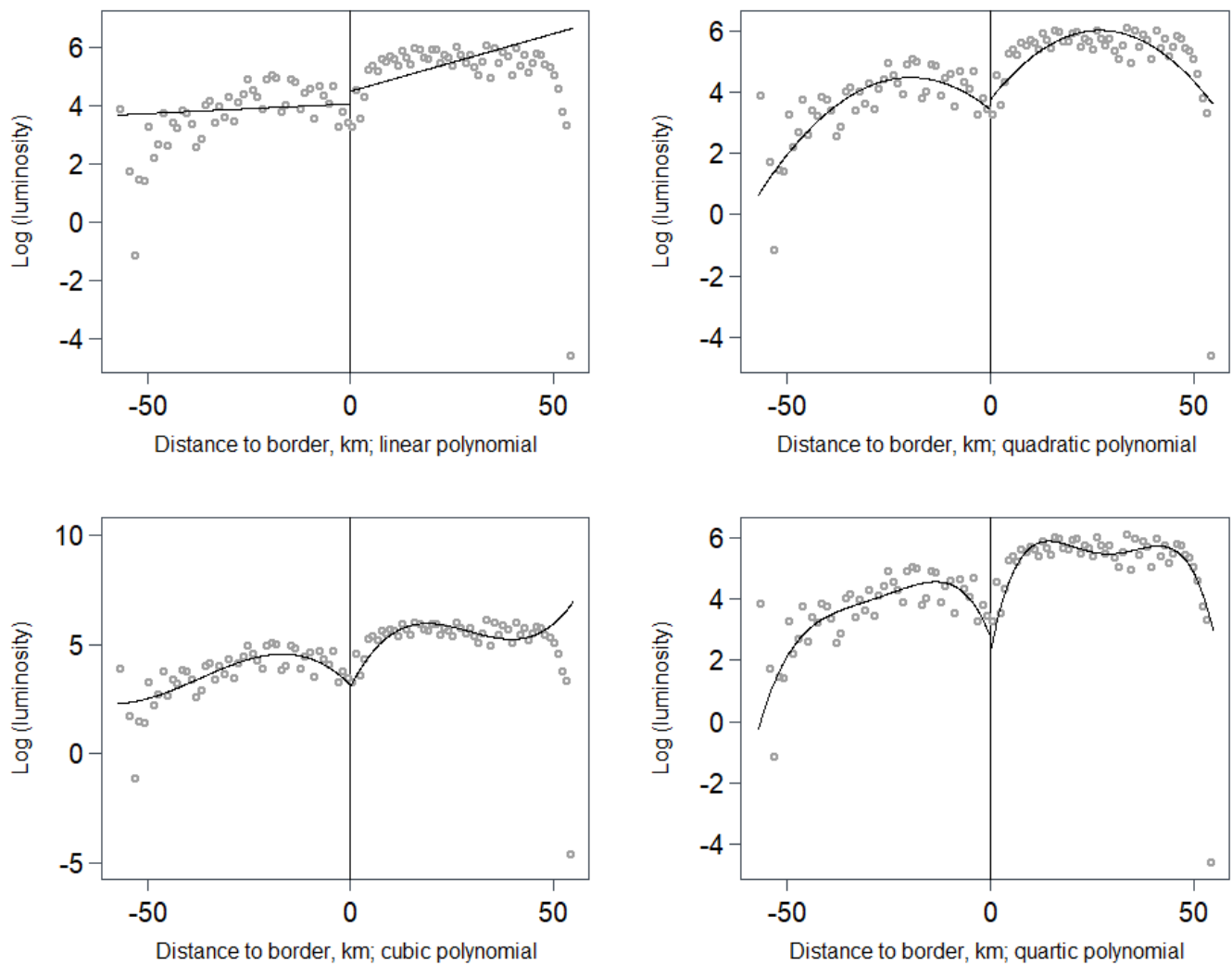

Figure 13. Regression discontinuity plots, Guangxi-Guangdong border.

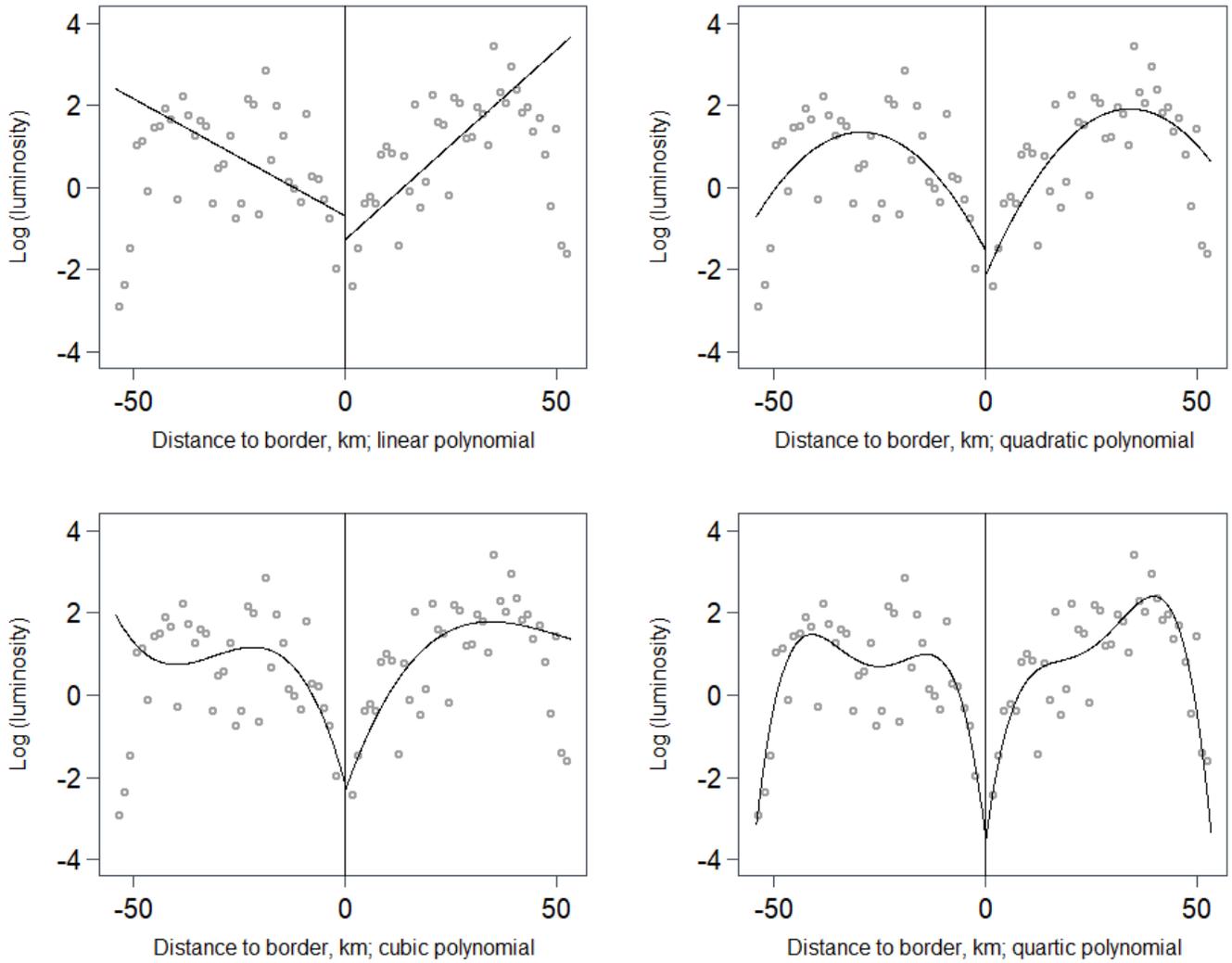

Table 1. Frequency distribution of dialect groups, bordering cells

| Group | Freq. | Percent | Cum. Pct. |
|---|---|---|---|
| Gan | 3,578 | 3.32 | 3.32 |
| Hakka | 4,143 | 3.84 | 7.16 |
| Jin | 10,201 | 9.46 | 16.63 |
| Mandarin | 72,353 | 67.13 | 83.76 |
| Mongolian | 6,141 | 5.70 | 89.46 |
| Tibeto-Burman | 7,562 | 7.02 | 96.47 |
| Wu | 2,544 | 2.36 | 98.83 |
| Yue (Cantonese) | 1,259 | 1.17 | 100.00 |
| $N$ | 107,781 | | |

Table 2. Control RDD regressions: Elevation, linear polynomial

| DV: Elevation | (1)<br>Anhui-Henan | (2)<br>Anhui-Hubei | (3)<br>Anhui-Jiangsu | (4)<br>Anhui-Jiangxi | (5)<br>Anhui-Shandong |
|---|---|---|---|---|---|
| High quality | -9.12<br>(23.236) | 14.903<br>(119.800) | -1.514<br>(4.540) | 18.418<br>(89.869) | -1.984<br>(1.626) |
| Bandwidth | 33.197 | 22.565 | 39.114 | 19.063 | 31.808 |
| Effective observations | 1207 | 287 | 1570 | 365 | 43 |
| Observations | 2,519 | 969 | 2,977 | 1,174 | 163 |

| DV: Elevation | (1)<br>Anhui-Zhejiang | (2)<br>Hubei-Chongqing | (3)<br>Hunan-Chongqing | (4)<br>Inner Mongolia-Gansu | (5)<br>Ningxia-Gansu |
|---|---|---|---|---|---|
| High quality | -22.217<br>(62.582) | 21.364<br>(100.000) | -144.000<br>(98.793) | -18.885<br>(32.194) | -13.084<br>(38.921) |
| Bandwidth | 28.736 | 21.957 | 22.666 | 21.494 | 23.645 |
| Effective observations | 454 | 659 | 119 | 1907 | 1026 |
| Observations | 1,233 | 2,180 | 459 | 6,287 | 2,910 |

| DV: Elevation | (1)<br>Sichuan-Gansu | (2)<br>Xinjiang-Gansu | (3)<br>Fujian-Guangdong | (4)<br>Guangxi-Guangdong | (5)<br>Hunan-Guangdong |
|---|---|---|---|---|---|
| High quality | -123.83<br>(157.830) | -76.074<br>(97.363) | -19.328<br>(77.538) | 15.923<br>(59.374) | -64.156<br>(118.320) |
| Bandwidth | 20.954 | 21.886 | 21.091 | 18.914 | 17.656 |
| Effective observations | 772 | 715 | 254 | 522 | 321 |
| Observations | 2,754 | 2,698 | 965 | 2,036 | 1,234 |

| DV: Elevation | (1)<br>Jiangxi-Guangdong | (2)<br>Hubei-Hunan | (3)<br>Jiangxi-Hunan | (4)<br>Liaoning-Jilin | (5)<br>Inner Mongolia-Liaoning |
|---|---|---|---|---|---|
| High quality | -53.313<br>(55.879) | -45.295<br>(76.076) | -6.1493<br>(93.705) | 11.771<br>(28.050) | -6.823<br>(33.392) |
| Bandwidth | 18.190 | 27.122 | 22.905 | 33.883 | 26.672 |
| Effective observations | 417 | 995 | 663 | 919 | 1036 |
| Observations | 1,543 | 2,801 | 2,002 | 2,095 | 2,922 |

| DV: Elevation | (1)<br>Hubei-Shaanxi | (2)<br>Shaanxi-Shanxi |
|---|---|---|
| High quality | 59.221<br>(123.02) | -18.175<br>(42.421) |
| Bandwidth | 16.544 | 21.318 |
| Effective observations | 403 | 695 |
| Observations | 1,519 | 2,694 |

Notes: Estimates derived from local linear regressions. Calculated via Stata's "rdrobust", fitting a linear polynomial model on either side of cutoff. Observations refers to number of cells in buffer on either side of border, and effective observations those used in regression, reflecting mean squared error-optimal bandwidth (in kilometres), also noted in the table. rdrobust calculates these bias-corrected bandwidths using companion command, "rdbwselect". Both parameter of interest and bandwidth are bias corrected, variance being estimated using three nearest neighbours. Locally-weighted averages computed using triangular kernel. Standard errors in parentheses. *, **, and *** indicate statistical significance at the 10, 5 and 1 percent levels, respectively.

Table 3. Control RDD regressions: Precipitation, linear polynomial

| DV: Elevation | (1)<br>Anhui-Henan | (2)<br>Anhui-Hubei | (3)<br>Anhui-Jiangsu | (4)<br>Anhui-Jiangxi | (5)<br>Anhui-Shandong |
|---|---|---|---|---|---|
| High quality | -11,032** | 4,341 | 652.18 | 9,035.7 | -3,124.4 |
|  | (5027.6) | (9591.4) | (3,177) | (10,191) | (7,981.5) |
| Bandwidth | 11.404 | 14.219 | 11.703 | 12.199 | 19.755 |
| Effective observations | 464 | 218 | 688 | 286 | 33 |
| Observations | 2,542 | 981 | 3,002 | 1,187 | 168 |
| DV: Elevation | (1)<br>Anhui-Zhejiang | (2)<br>Hubei-Chongqing | (3)<br>Hunan-Chongqing | (4)<br>Inner Mongolia-Gansu | (5)<br>Ningxia-Gansu |
| High quality | -9,618.5 | -7,037.4 | 4,825.9 | 96.054 | 1,341.6 |
|  | (8,793.7) | (7,086.8) | (12,679) | (173.33) | (865.04) |
| Bandwidth | 14.473 | 12.252 | 16.140 | 10.809 | 11.182 |
| Effective observations | 271 | 440 | 89 | 1114 | 634 |
| Observations | 1,244 | 2,203 | 467 | 6,340 | 2,936 |
| DV: Elevation | (1)<br>Sichuan-Gansu | (2)<br>Xinjiang-Gansu | (3)<br>Fujian-Guangdong | (4)<br>Guangxi-Guangdong | (5)<br>Hunan-Guangdong |
| High quality | 2,279.1 | 328.94* | -11,987 | 2,698.6 | 3,136 |
|  | (3,016.4) | (170.92) | (10,763) | (7,772.9) | (6,681.4) |
| Bandwidth | 11.368 | 13.332 | 13.772 | 11.720 | 12.826 |
| Effective observations | 539 | 450 | 190 | 401 | 302 |
| Observations | 2,772 | 2,707 | 977 | 2,062 | 1,247 |
| DV: Elevation | (1)<br>Jiangxi-Guangdong | (2)<br>Hubei-Hunan | (3)<br>Jiangxi-Hunan | (4)<br>Liaoning-Jilin | (5)<br>Inner Mongolia-Liaoning |
| High quality | 4,847.1 | -983.79 | -5,067.3 | 1,521.5 | -985.25 |
|  | (8,516.1) | (5,366.8) | (7,680.4) | (1,490.7) | (812.01) |
| Bandwidth | 12.818 | 12.553 | 11.798 | 11.668 | 10.008 |
| Effective observations | 356 | 618 | 411 | 438 | 556 |
| Observations | 1,552 | 2,825 | 2,025 | 2,107 | 2,948 |
| DV: Elevation | (1)<br>Hubei-Shaanxi | (2)<br>Shaanxi-Shanxi | | | |
| High quality | 1,224.9 | -292.04 | | | |
|  | (4,123.3) | (714.51) | | | |
| Bandwidth | 12.375 | 9.643 | | | |
| Effective observations | 368 | 462 | | | |
| Observations | 1,532 | 2,722 | | | |

Notes: Estimates derived from local linear regressions. Calculated via Stata's "rdrobust" command that fits a linear polynomial model on either side of cutoff. Observations refers to number of cells in buffer on either side of border, and effective observations those used in regression, reflecting mean squared error-optimal bandwidth (in kilometres), also noted in the table. rdrobust calculates these bias-corrected bandwidths using companion command, "rdbwselect". Both parameter of interest and bandwidth are bias corrected, variance being estimated using three nearest neighbours. Locally-weighted averages computed using triangular kernel. Standard errors in parentheses. *, **, and *** indicate statistical significance at the 10, 5 and 1 percent levels, respectively.

Table 4. Control RDD regressions: Road density, linear polynomial

| DV: Dist_road | (1)<br>Anhui-Henan | (2)<br>Anhui-Hubei | (3)<br>Anhui-Jiangsu | (4)<br>Anhui-Jiangxi | (5)<br>Anhui-Shandong |
|---|---|---|---|---|---|
| High quality | -0.103 | -0.461 | -0.671 | 0.216 | -1.201 |
| | (0.609) | (1.331) | (0.749) | (1.586) | (2.581) |
| Bandwidth | 21.879 | 22.838 | 22.761 | 19.616 | 25.718 |
| Effective observations | 755 | 289 | 978 | 376 | 34 |
| Observations | 2,542 | 981 | 3,002 | 1,187 | 168 |

| DV: Dist_road | (1)<br>Anhui-Zhejiang | (2)<br>Hubei-Chongqing | (3)<br>Hunan-Chongqing | (4)<br>Inner Mongolia-<br>Gansu | (5)<br>Ningxia-Gansu |
|---|---|---|---|---|---|
| High quality | -0.272 | -0.403 | -2.289 | -0.519 | 0.910 |
| | (0.827) | (1.800) | (2.618) | (1.362) | (0.967) |
| Bandwidth | 21.182 | 21.922 | 18.233 | 26.919 | 22.064 |
| Effective observations | 340 | 674 | 85 | 2122 | 998 |
| Observations | 1,244 | 2,203 | 467 | 6,340 | 2,936 |

| DV: Dist_road | (1)<br>Sichuan-Gansu | (2)<br>Xinjiang-Gansu | (3)<br>Fujian-Guangdong | (4)<br>Guangxi-Guangdong | (5)<br>Hunan-Guangdong |
|---|---|---|---|---|---|
| High quality | -0.121 | 4.318 | 1.036 | 1.126 | -0.722 |
| | (1.205) | (5.054) | (1.374) | (0.963) | (1.221) |
| Bandwidth | 25.466 | 21.446 | 19.218 | 15.953 | 23.031 |
| Effective observations | 896 | 725 | 221 | 425 | 423 |
| Observations | 2,772 | 2,707 | 977 | 2,062 | 1,247 |

| DV: Dist_road | (1)<br>Jiangxi-Guangdong | (2)<br>Hubei-Hunan | (3)<br>Jiangxi-Hunan | (4)<br>Liaoning-Jilin | (5)<br>Inner Mongolia-<br>Liaoning |
|---|---|---|---|---|---|
| High quality | 0.400 | 4.084* | -0.313 | -0.088 | 0.233 |
| | (0.732) | (2.130) | (0.702) | (0.573) | (0.417) |
| Bandwidth | 31.121 | 14.910 | 26.539 | 22.537 | 28.362 |
| Effective observations | 622 | 533 | 662 | 690 | 1093 |
| Observations | 1,552 | 2,825 | 2,025 | 2,107 | 2,948 |

| DV: Dist_road | (1)<br>Hubei-Shaanxi | (2)<br>Shaanxi-Shanxi |
|---|---|---|
| High quality | -1.631 | 0.399 |
| | (2.292) | (1.325) |
| Bandwidth | 17.675 | 27.680 |
| Effective observations | 402 | 931 |
| Observations | 1,532 | 2,722 |

Notes: Estimates derived from local linear regressions. Calculated via Stata's "rdrobust" command that fits a linear polynomial model on either side of cutoff. Observations refers to number of cells in buffer on either side of border, and effective observations those used in regression, reflecting mean squared error-optimal bandwidth (in kilometres), also noted in the table. rdrobust calculates these bias-corrected bandwidths using companion command, "rdbwselect". Both parameter of interest and bandwidth are bias corrected, variance being estimated using three nearest neighbours. Locally-weighted averages computed using triangular kernel. Standard errors in parentheses. *, **, and *** indicate statistical significance at the 10, 5 and 1 percent levels, respectively.

Table 5. Control RDD regressions: Population, linear polynomial

| DV: Population | (1) Anhui-Henan | (2) Anhui-Hubei | (3) Anhui-Jiangsu | (4) Anhui-Jiangxi | (5) Anhui-Shandong |
|---|---|---|---|---|---|
| High quality | 0.120 | 0.087 | -0.228 | 0.033 | 1.002* |
| | (0.115) | (0.164) | (0.163) | (0.283) | (0.562) |
| Bandwidth | 31.424 | 31.212 | 20.650 | 16.066 | 22.077 |
| Effective observations | 959 | 276 | 757 | 233 | 20 |
| Observations | 2,307 | 870 | 2,712 | 1,041 | 112 |

| DV: Population | (1) Anhui-Zhejiang | (2) Hubei-Chongqing | (3) Hunan-Chongqing | (4) Inner Mongolia-Gansu | (5) Ningxia-Gansu |
|---|---|---|---|---|---|
| High quality | 0.207 | -0.089 | 0.066 | 0.600 | -0.028 |
| | (0.252) | (0.098) | (0.387) | (0.367) | (0.119) |
| Bandwidth | 18.749 | 26.538 | 19.511 | 19.258 | 29.339 |
| Effective observations | 331 | 579 | 70 | 1247 | 1000 |
| Observations | 1,087 | 1,968 | 360 | 5,849 | 2,644 |

| DV: Population | (1) Sichuan-Gansu | (2) Xinjiang-Gansu | (3) Fujian-Guangdong | (4) Guangxi-Guangdong | (5) Hunan-Guangdong |
|---|---|---|---|---|---|
| High quality | -0.031 | -0.455 | -0.044 | -0.023 | 0.178 |
| | (0.196) | (0.288) | (0.230) | (0.171) | (0.215) |
| Bandwidth | 26.062 | 33.693 | 25.996 | 20.691 | 18.967 |
| Effective observations | 823 | 933 | 255 | 471 | 263 |
| Observations | 2,511 | 2,518 | 840 | 1,853 | 1,096 |

| DV: Population | (1) Jiangxi-Guangdong | (2) Hubei-Hunan | (3) Jiangxi-Hunan | (4) Liaoning-Jilin | (5) Inner Mongolia-Liaoning |
|---|---|---|---|---|---|
| High quality | 0.128 | 0.150 | 0.123 | 0.165 | -0.050 |
| | (0.145) | (0.132) | (0.207) | (0.190) | (0.123) |
| Bandwidth | 21.945 | 33.779 | 21.455 | 21.529 | 17.606 |
| Effective observations | 415 | 924 | 477 | 481 | 564 |
| Observations | 1,386 | 2,545 | 1,829 | 1,895 | 2,680 |

| DV: Population | (1) Hubei-Shaanxi | (2) Shaanxi-Shanxi |
|---|---|---|
| High quality | 0.401* | -0.309 |
| | (0.219) | (0.206) |
| Bandwidth | 14.462 | 23.077 |
| Effective observations | 258 | 643 |
| Observations | 1,377 | 2,474 |

Notes: Estimates derived from local linear regressions. Calculated via Stata's "rdrobust" command that fits a linear polynomial model on either side of cutoff. Observations refers to number of cells in buffer on either side of border, and effective observations those used in regression, reflecting mean squared error-optimal bandwidth (in kilometres), also noted in the table. rdrobust calculates these bias-corrected bandwidths using companion command, "rdbwselect". Both parameter of interest and bandwidth are bias corrected, variance being estimated using three nearest neighbours. Locally-weighted averages computed using triangular kernel. Standard errors in parentheses. *, **, and *** indicate statistical significance at the 10, 5 and 1 percent levels, respectively.

Table 6. RDD estimates with dialect group fixed effects

| Dependent variable | (1) Luminosity (linear) | (2) Luminosity (quadratic) | (3) Luminosity per person (linear) | (4) Luminosity per person (quadratic) | (5) Lit (linear) | (6) Lit (quadratic) |
|---|---|---|---|---|---|---|
| High quality institutions | -0.0168 | -0.0527 | -0.0292 | -0.0538 | -0.00809 | -0.0251 |
|  | (0.241) | (0.274) | (0.0537) | (0.0555) | (0.0287) | (0.0299) |
| Observations | 107,781 | 107,781 | 98,831 | 98,831 | 107,781 | 107,781 |
| Effective obs., left of cutoff | 9454 | 15429 | 7560 | 15434 | 9824 | 17269 |
| Effective obs., right of cutoff | 8724 | 14262 | 7015 | 14278 | 9091 | 16001 |
| Bandwidth | 5.369 | 10.53 | 5.953 | 12.99 | 5.678 | 12.19 |
| Bandwidth bias | 11.94 | 15.93 | 14.36 | 21.94 | 13.09 | 18.06 |
| Kernel | Triangular | Triangular | Triangular | Triangular | Triangular | Triangular |

Notes: Estimates derived from local linear regressions. Calculated via Stata's "rdrobust" command that fits a linear polynomial model on either side of cutoff. Observations refers to number of cells in buffer on either side of border, and effective observations those used in regression, reflecting mean squared error-optimal bandwidth (in kilometres), also noted in the table. rdrobust calculates these bias-corrected bandwidths using companion command, "rdbwselect". Both parameter of interest and bandwidth are bias corrected, variance being estimated using three nearest neighbours. Locally-weighted averages computed using triangular kernel. Standard errors in parentheses. *, **, and *** indicate statistical significance at the 10, 5 and 1 percent levels, respectively.

Table 7. RDD estimates with dialect group fixed effects and control variables

| Dependent variable | (1) Luminosity (linear) | (2) Luminosity (quadratic) | (3) Luminosity per person (linear) | (4) Luminosity per person (quadratic) | (5) Lit (linear) | (6) Lit (quadratic) |
|---|---|---|---|---|---|---|
| High quality institutions | -0.0188 | -0.399* | -0.0322 | -0.0741* | -0.00245 | -0.0452 |
|  | (0.296) | (0.233) | (0.0485) | (0.0400) | (0.0364) | (0.0276) |
| Observations | 96,551 | 96,551 | 96,551 | 96,551 | 96,551 | 96,551 |
| Effective obs., left of cutoff | 6704 | 15860 | 7411 | 17876 | 6461 | 15353 |
| Effective obs., right of cutoff | 6474 | 15306 | 7205 | 17306 | 6238 | 14823 |
| Bandwidth | 5.531 | 14.11 | 6.121 | 16.13 | 5.314 | 13.62 |
| Bandwidth bias | 15.61 | 25.73 | 16.52 | 28.94 | 15.43 | 24.92 |
| kernel | Triangular | Triangular | Triangular | Triangular | Triangular | Triangular |

Notes: Estimates derived from local linear regressions. Calculated via Stata's "rdrobust" command that fits a linear polynomial model on either side of cutoff. Observations refers to number of cells in buffer on either side of border, and effective observations those used in regression, reflecting mean squared error-optimal bandwidth (in kilometres), also noted in the table. rdrobust calculates these bias-corrected bandwidths using companion command, "rdbwselect". Both parameter of interest and bandwidth are bias corrected, variance being estimated using three nearest neighbours. Locally-weighted averages computed using triangular kernel. Standard errors in parentheses. *, **, and *** indicate statistical significance at the 10, 5 and 1 percent levels, respectively.

Table 8. Regression discontinuity estimates, local linear regressions. Dependent variable: *luminosity*

| DV: Luminosity | (1) Anhui-Henan | (2) Anhui-Hubei | (3) Anhui-Jiangsu | (4) Anhui-Jiangxi | (5) Anhui-Shandong |
|---|---|---|---|---|---|
| High quality | 1.153 | -2.285 | 0.024 | 0.033 | Insufficient obs. |
| | (1.030) | (1.504) | (0.652) | (1.082) | |
| Bandwidth | 16.664 | 16.175 | 15.552 | 22.159 | |
| Effective observations | 500 | 186 | 569 | 328 | |
| Observations | 2,306 | 868 | 2,711 | 1,040 | |

| DV: Luminosity | (1) Anhui-Zhejiang | (2) Hubei-Chongqing | (3) Hunan-Chongqing | (4) Inner Mongolia-Gansu | (5) Ningxia-Gansu |
|---|---|---|---|---|---|
| High quality | -3.025** | 0.613 | -2.074 | -0.188 | 0.646 |
| | (1.369) | (0.533) | (2.422) | (0.186) | (0.742) |
| Bandwidth | 20.569 | 26.060 | 16.553 | 17.562 | 19.926 |
| Effective observations | 219 | 581 | 62 | 1235 | 674 |
| Observations | 1,086 | 1,966 | 360 | 5,847 | 2,643 |

| DV: Luminosity | (1) Sichuan-Gansu | (2) Xinjiang-Gansu | (3) Fujian-Guangdong | (4) Guangxi-Guangdong | (5) Hunan-Guangdong |
|---|---|---|---|---|---|
| High quality | -0.063 | -0.048 | -1.046 | -0.580 | -0.308 |
| | (0.471) | (0.370) | (1.080) | (0.914) | (0.997) |
| Bandwidth | 20.073 | 22.745 | 31.756 | 23.448 | 24.598 |
| Effective observations | 601 | 667 | 312 | 539 | 329 |
| Observations | 2,509 | 2,516 | 839 | 1,852 | 1,095 |

| DV: Luminosity | (1) Jiangxi-Guangdong | (2) Hubei-Hunan | (3) Jiangxi-Hunan | (4) Liaoning-Jilin | (5) Inner Mongolia-Liaoning |
|---|---|---|---|---|---|
| High quality | -1.077 | -2.23* | -0.806 | 0.169 | 0.390 |
| | (0.989) | (1.286) | (0.912) | (0.852) | (0.995) |
| Bandwidth | 20.813 | 16.817 | 22.333 | 25.743 | 19.737 |
| Effective observations | 368 | 484 | 499 | 602 | 614 |
| Observations | 1,386 | 2,542 | 1,828 | 1,894 | 2,680 |

| DV: Luminosity | (1) Hubei-Shaanxi | (2) Shaanxi-Shanxi |
|---|---|---|
| High quality | -0.990 | 0.547 |
| | (0.874) | (1.062) |
| Bandwidth | 18.923 | 20.856 |
| Effective observations | 339 | 537 |
| Observations | 1,376 | 2,471 |

Notes: Estimates derived from local linear regressions. Calculated via Stata's "rdrobust" command that fits a linear model on either side of cutoff. Estimates and bandwidths are adjusted for covariates noted in main text. Observations refers to number of cells in buffer on either side of border, and effective observations those used in regression, reflecting mean squared error-optimal bandwidth (in kilometres), also noted in the table. rdrobust calculates these bias-corrected bandwidths using companion command, "rdbwselect". Both parameter of interest and bandwidth are bias corrected, variance being estimated using three nearest neighbours. Locally-weighted averages computed using triangular kernel. Standard errors in parentheses. *, **, and *** indicate statistical significance at the 10, 5 and 1 percent levels, respectively.

Table 9. Regression discontinuity estimates, local linear regressions (quadratic polynomial). Dependent variable: *luminosity*

| DV: Luminosity | (1)<br>Anhui-Henan | (2)<br>Anhui-Hubei | (3)<br>Anhui-Jiangsu | (4)<br>Anhui-Jiangxi | (5)<br>Anhui-Shandong |
|---|---|---|---|---|---|
| High quality | 1.534 | -2.567 | 0.061 | 0.713 | -2.822* |
| | (1.045) | (2.034) | (0.768) | (1.835) | (1.659) |
| Bandwidth | 27.456 | 21.593 | 23.170 | 25.855 | 18.292 |
| Effective observations | 920 | 257 | 949 | 401 | 19 |
| Observations | 2306 | 868 | 2711 | 1040 | 112 |

| DV: Luminosity | (1)<br>Anhui-Zhejiang | (2)<br>Hubei-Chongqing | (3)<br>Hunan-Chongqing | (4)<br>Inner Mongolia-Gansu | (5)<br>Ningxia-Gansu |
|---|---|---|---|---|---|
| High quality | -3.805* | -0.409 | -4.166 | -0.208 | 0.044 |
| | (2.025) | (0.823) | (3.922) | (0.252) | (1.210) |
| Bandwidth | 23.189 | 31.004 | 21.879 | 20.644 | 22.951 |
| Effective observations | 317 | 744 | 84 | 1508 | 874 |
| Observations | 1,086 | 1,966 | 360 | 5,847 | 2,643 |

| DV: Luminosity | (1)<br>Sichuan-Gansu | (2)<br>Xinjiang-Gansu | (3)<br>Fujian-Guangdong | (4)<br>Guangxi-Guangdong | (5)<br>Hunan-Guangdong |
|---|---|---|---|---|---|
| High quality | -0.264 | -0.044 | -0.839 | -0.630 | -0.419 |
| | (0.497) | (0.462) | (1.662) | (1.277) | (1.641) |
| Bandwidth | 35.714 | 32.074 | 34.848 | 30.238 | 25.932 |
| Effective observations | 1049 | 984 | 372 | 720 | 394 |
| Observations | 2,509 | 2,516 | 839 | 1,852 | 1,095 |

| DV: Luminosity | (1)<br>Jiangxi-Guangdong | (2)<br>Hubei-Hunan | (3)<br>Jiangxi-Hunan | (4)<br>Liaoning-Jilin | (5)<br>Inner Mongolia-Liaoning |
|---|---|---|---|---|---|
| High quality | -1.702 | -2.504 | -1.389 | 2.976 | 0.283 |
| | (1.908) | (1.777) | (1.685) | (1.231) | (1.319) |
| Bandwidth | 22.276 | 22.318 | 24.540 | 21.565 | 25.852 |
| Effective observations | 444 | 735 | 602 | 516 | 947 |
| Observations | 1,386 | 2,542 | 1,828 | 1,894 | 2,680 |

| DV: Luminosity | (1)<br>Hubei-Shaanxi | (2)<br>Shaanxi-Shanxi |
|---|---|---|
| High quality | -1.211 | 0.956 |
| | (1.239) | (1.476) |
| Bandwidth | 24.544 | 26.078 |
| Effective observations | 493 | 785 |
| Observations | 1,376 | 2,471 |

Notes: Estimates derived from local linear regressions. Calculated via Stata's "rdrobust" command using quadratic polynomial that varies on either side of cutoff. Estimates and bandwidths are adjusted for covariates noted in main text. Observations refers to number of cells in buffer on either side of border, and effective observations those used in regression, reflecting mean squared error-optimal bandwidth (in kilometres), also noted in the table. rdrobust calculates these bias-corrected bandwidths using companion command, "rdbwselect". Both parameter of interest and bandwidth are bias corrected, variance being estimated using three nearest neighbours. Locally-weighted averages computed using triangular kernel. Standard errors in parentheses. *, **, and *** indicate statistical significance at the 10, 5 and 1 percent levels, respectively.

Table 10. Regression discontinuity estimates, local linear regressions.  Dependent variable: *lit*

| | (1) | (2) | (3) | (4) | (5) |
|---|---|---|---|---|---|
| DV: Luminosity | Anhui-Henan | Anhui-Hubei | Anhui-Jiangsu | Anhui-Jiangxi | Anhui-Shandong |
| High quality | 0.099 | -0.333* | 0.028 | 0.060 | - |
| | (0.100) | (0.192) | (0.069) | (0.131) | Multicollinearity in covs |
| Bandwidth | 17.887 | 15.624 | 15.807 | 21.328 | |
| Effective observations | 548 | 169 | 597 | 312 | |
| Observations | 2306 | 868 | 2711 | 1040 | |
| | (1) | (2) | (3) | (4) | (5) |
| DV: Luminosity | Anhui-Zhejiang | Hubei-Chongqing | Hunan-Chongqing | Inner Mongolia-Gansu | Ningxia-Gansu |
| High quality | -0.228 | 0.102 | -0.388 | -0.031 | 0.054 |
| | (0.147) | (0.065) | (0.287) | (0.026) | (0.903) |
| Bandwidth | 23.497 | 27.173 | 16.352 | 16.459 | 20.850 |
| Effective observations | 241 | 583 | 62 | 1191 | 709 |
| Observations | 1,086 | 1,966 | 360 | 5,847 | 2,643 |
| | (1) | (2) | (3) | (4) | (5) |
| DV: Luminosity | Sichuan-Gansu | Xinjiang-Gansu | Fujian-Guangdong | Guangxi-Guangdong | Hunan-Guangdong |
| High quality | -0.025 | -0.020 | -0.059 | -0.063 | -0.041 |
| | (0.055) | (0.468) | (0.129) | (0.103) | (0.111) |
| Bandwidth | 21.825 | 22.532 | 30.954 | 25.955 | 27.363 |
| Effective observations | 640 | 630 | 305 | 587 | 343 |
| Observations | 2,509 | 2,516 | 839 | 1,852 | 1,095 |
| | (1) | (2) | (3) | (4) | (5) |
| DV: Luminosity | Jiangxi-Guangdong | Hubei-Hunan | Jiangxi-Hunan | Liaoning-Jilin | Inner Mongolia-Liaoning |
| High quality | -0.150 | -0.311 | -0.098 | 0.359 | 0.055 |
| | (0.120) | (0.158) | (0.112) | (0.105) | (0.107) |
| Bandwidth | 20.714 | 16.886 | 21.643 | 24.582 | 21.140 |
| Effective observations | 360 | 461 | 491 | 583 | 674 |
| Observations | 1,386 | 2,542 | 1,828 | 1,894 | 2,680 |
| | (1) | (2) | | | |
| DV: Luminosity | Hubei-Shaanxi | Shaanxi-Shanxi | | | |
| High quality | -0.138 | 0.097 | | | |
| | (0.113) | (0.114) | | | |
| Bandwidth | 18.898 | 22.977 | | | |
| Effective observations | 339 | 590 | | | |
| Observations | 1,376 | 2,471 | | | |

Notes: Estimates derived from local linear regressions.  Calculated via Stata's "rdrobust" command using linear polynomial that varies on either side of cutoff.  Estimates and bandwidths are adjusted for covariates noted in main text.  Observations refers to number of cells in buffer on either side of border, and effective observations those used in regression, reflecting mean squared error-optimal bandwidth (in kilometres), also noted in the table.  rdrobust calculates these bias-corrected bandwidths using companion command, "rdbwselect".  Both parameter of interest and bandwidth are bias corrected, variance being estimated using three nearest neighbours.  Locally-weighted averages computed using triangular kernel.  Standard errors in parentheses.  *, **, and *** indicate statistical significance at the 10, 5 and 1 percent levels, respectively.

Table 11. Regression discontinuity estimates, local linear regressions (quadratic polynomial). Dependent variable: *lit*

| DV: Luminosity | (1) Anhui-Henan | (2) Anhui-Hubei | (3) Anhui-Jiangsu | (4) Anhui-Jiangxi | (5) Anhui-Shandong |
|---|---|---|---|---|---|
| High quality | 0.124 | -0.411 | 0.036 | 0.162 | - |
|  | (0.102) | (0.257) | (0.079) | (0.227) | Multicollinearity in covs |
| Bandwidth | 29.724 | 20.726 | 23.822 | 24.720 |  |
| Effective observations | 991 | 247 | 1006 | 386 |  |
| Observations | 2306 | 868 | 2711 | 1040 |  |

| DV: Luminosity | (1) Anhui-Zhejiang | (2) Hubei-Chongqing | (3) Hunan-Chongqing | (4) Inner Mongolia-Gansu | (5) Ningxia-Gansu |
|---|---|---|---|---|---|
| High quality | -0.476 | 0.002 | -0.681 | -0.032 | 0.084 |
|  | (0.259) | (0.098) | (0.451) | (0.034) | (0.101) |
| Bandwidth | 22.789 | 31.682 | 21.321 | 20.837 | 32.876 |
| Effective observations | 311 | 785 | 84 | 1560 | 1198 |
| Observations | 1,086 | 1,966 | 360 | 5,847 | 2,643 |

| DV: Luminosity | (1) Sichuan-Gansu | (2) Xinjiang-Gansu | (3) Fujian-Guangdong | (4) Guangxi-Guangdong | (5) Hunan-Guangdong |
|---|---|---|---|---|---|
| High quality | -0.032 | -0.031 | 0.010 | -0.073 | -0.044 |
|  | (0.066) | (0.054) | (0.206) | (0.146) | (0.162) |
| Bandwidth | 33.064 | 33.908 | 32.777 | 31.813 | 30.940 |
| Effective observations | 1006 | 1047 | 363 | 767 | 466 |
| Observations | 2,509 | 2,516 | 839 | 1,852 | 1,095 |

| DV: Luminosity | (1) Jiangxi-Guangdong | (2) Hubei-Hunan | (3) Jiangxi-Hunan | (4) Liaoning-Jilin | (5) Inner Mongolia-Liaoning |
|---|---|---|---|---|---|
| High quality | -0.225 | -0.424 | -0.180 | -0.328 | 0.059 |
|  | (0.231) | (0.221) | (0.205) | (0.261) | (0.151) |
| Bandwidth | 22.076 | 21.938 | 23.773 | 22.089 | 26.221 |
| Effective observations | 441 | 696 | 587 | 530 | 957 |
| Observations | 1,386 | 2,542 | 1,828 | 1,894 | 2,680 |

| DV: Luminosity | (1) Hubei-Shaanxi | (2) Shaanxi-Shanxi |
|---|---|---|
| High quality | -0.172 | 0.186 |
|  | (0.150) | (0.189) |
| Bandwidth | 25.558 | 25.089 |
| Effective observations | 526 | 726 |
| Observations | 1,376 | 2,471 |

Notes: Estimates derived from local linear regressions. Calculated via Stata's "rdrobust" command using quadratic polynomial that varies on either side of cutoff. Estimates and bandwidths are adjusted for covariates noted in main text. Observations refers to number of cells in buffer on either side of border, and effective observations those used in regression, reflecting mean squared error-optimal bandwidth (in kilometres), also noted in the table. rdrobust calculates these bias-corrected bandwidths using companion command, "rdbwselect". Both parameter of interest and bandwidth are bias corrected, variance being estimated using three nearest neighbours. Locally-weighted averages computed using triangular kernel. Standard errors in parentheses. *, **, and *** indicate statistical significance at the 10, 5 and 1 percent levels, respectively.

Table 12. Data sources

| Variable | Operationalization | Source |
|---|---|---|
| *Luminosity* | Log of sum of digital numbers within cell | National Oceanic and Atmospheric Association: National Centers for Environmental Information. http://ngdc.noaa.gov/eog/dmsp.html. |
| *Lit* | Dummy =1 if *Luminosity* >0 | As above. |
| *Population* | Log of total population within cell, 2005 | Gridded Population of the World, Version 4 (CIESIN, 2016). |
| *Elevation* | Metres above sea level, cell/province mean | China Historical Geographic Information System (CHGIS), Fairbank Center (2016). |
| *Ruggedness* | Standard deviation in *Elevation*, cell/province mean | China Historical Geographic Information System (CHGIS), Fairbank Center (2016). |
| *Road density* | Average distance to nearest road among pixels within cell | Global Roads Open Access Dataset, Version 1 (CIESIN, 2013). |
| *Precipitation* | Sum of annual precipitation within cell/province, averaged of years 1970-2000. | WorldClim – Global Climate Data (Fick and Hijmans, 2017). |
| *Distance to road* | Geodesic distance from cell centroid to nearest road. | Global Roads Open Access Dataset, Version 1 (CIESIN, 2013). |
| *Percent private* | Percentage of employees in prefecture employed in urban private enterprises or self-employed. | China City Statistical Yearbooks. Accessed via China Data Center, University of Michigan. |